\documentclass[twocolumn,preprintnumbers,amsmath,amssymb]{revtex4}
\usepackage{graphicx}% Include figure files
\usepackage{dcolumn}% Align table columns on decimal point
\usepackage{bm}% bold math
\usepackage{color}

\begin{document}

%\preprint{APS/123-QED}

%\title{Pros and cons of receptor clustering}
%\title{Receptor clustering: delicate trade-off between beneficial signal and detrimental noise amplification}
%\title{Optimal signaling by receptor clusters determined by intrinsic and extrinsic noise}
\title{Optimal receptor-cluster size determined by intrinsic and extrinsic noise}
%\title{Optimal receptor-cluster size from trade-off between signal and noise amplification}

\author{Gerardo Aquino, Diana Clausznitzer, Sylvain Tollis, Robert G. Endres}
 \affiliation{
   {Division of Molecular Biosciences, Imperial College London, 
    London SW7 2AZ, United Kingdom}\\
   {Centre for Integrative Systems Biology at Imperial College, Imperial College London, 
    London SW7 2AZ, United Kingdom}}

\date{\today}

\begin{abstract}

Biological cells sense external chemical stimuli in their environment using cell-surface receptors.
To increase the sensitivity of sensing, receptors often cluster. This process occurs most noticeably 
in bacterial chemotaxis, a paradigm for sensing and signaling in general. While amplification of weak 
stimuli is useful in the absence of noise, its usefulness is less clear in the presence of extrinsic input 
noise and intrinsic signaling noise. Here, exemplified in a bacterial chemotaxis system, we combine the 
allosteric Monod-Wyman-Changeux model for signal amplification by receptor complexes with calculations 
of noise to study their interconnectedness. Importantly, we calculate the signal-to-noise ratio, describing 
the balance of beneficial and detrimental effects of clustering for the cell. Interestingly, we find 
that there is no advantage for the cell to build receptor complexes for noisy input stimuli
in the absence of intrinsic signaling noise. However, with intrinsic noise, an 
optimal complex size arises in line with estimates of the size of chemoreceptor complexes in bacteria 
and protein aggregates in lipid rafts of eukaryotic cells.

\end{abstract}

%\pacs{87.10.Mn, 87.15.kp, 87.16.dj}

%\keywords{Suggested keywords}%Use showkeys class option if keyword
                              %display desired
\maketitle

\section{\label{sec1} Introduction}
Biological cells can sense and respond to various chemicals in their environment. However, the 
precision with which a cell can measure and internally evaluate the concentration of a specific
ligand molecule is negatively affected by many sources of noise \cite{raj08,sha08}. 
There is external input noise (extrinsic noise) from the random arrival of ligand molecules at the cell-surface 
receptors by diffusion \cite{ber77,tka08,end08b}, as well as various sources of intracellular signaling noise 
(intrinsic noise) due, {\it e.g.}, to receptor dynamics, adaptation, and signal transduction \cite{yu08}, all relying on 
random chemical events. Nonetheless several biological examples 
exist in which measurements are performed with surprisingly high sensitivity. In bacterial chemotaxis, 
for instance, the bacterium {\it Escherichia coli} can respond to changes in 
concentration as low as 3.2 nM \cite{man03}, corresponding to only three molecules in the cell volume. 
High sensitivity is observed also in spatial sensing by single cell eukaryotic organisms, such as during aggregation
of the social amoeba {\it Dictyostelium discoideum} \cite{haa07} and during mating of 
{\it Saccharomyces cerevisiae} (budding yeast) \cite{man04}.
Furthermore, axon growth cones of neurons respond to an estimated change in concentration of about one molecule in the 
volume of the growth cone \cite{mor09}, and T cells of our immune system 
respond to a single peptide-major histocompatibility complex 
on a target cell \cite{syk96}. How can this sensitivity be understood despite
the various sources of noise?

The best characterized signal-transduction pathway is the bacterial chemotaxis pathway, allowing cells to swim
to sources of nutrients such as sugars and amino acids, and away from toxins \cite{ber99}. 
Cells are equipped with different receptor types with Tar among the most abundant receptors (hundreds to 
thousands of copies per cell). Tar specifically binds aspartate (or its non-metabolizable analogue MeAsp). 
An increase in ligand concentration, as occurring, e.g., when the cell swims towards 
the source of an attractant, inhibits receptor signaling activity and keeps the cell on course. In contrast, 
a decrease in attractant concentration, as occurring, e.g., when the cell swims in the wrong direction, increases
receptor signaling activity. This enhances the probability for the cell to randomly find a new and hopefully better 
direction of swimming. Cells are further equipped with an adaptation mechanism, which allows 
them to sense changes in ligand concentration over a wide range of background concentration.
Specifically, cells adapt their signaling activity by receptor methylation and demethylation. 
Methylation by enzyme CheR increases the receptor signaling activity, while demethylation by enzyme CheB
decreases receptor activity. 

\begin{figure*}[t]
\includegraphics[width=17cm,angle=0]{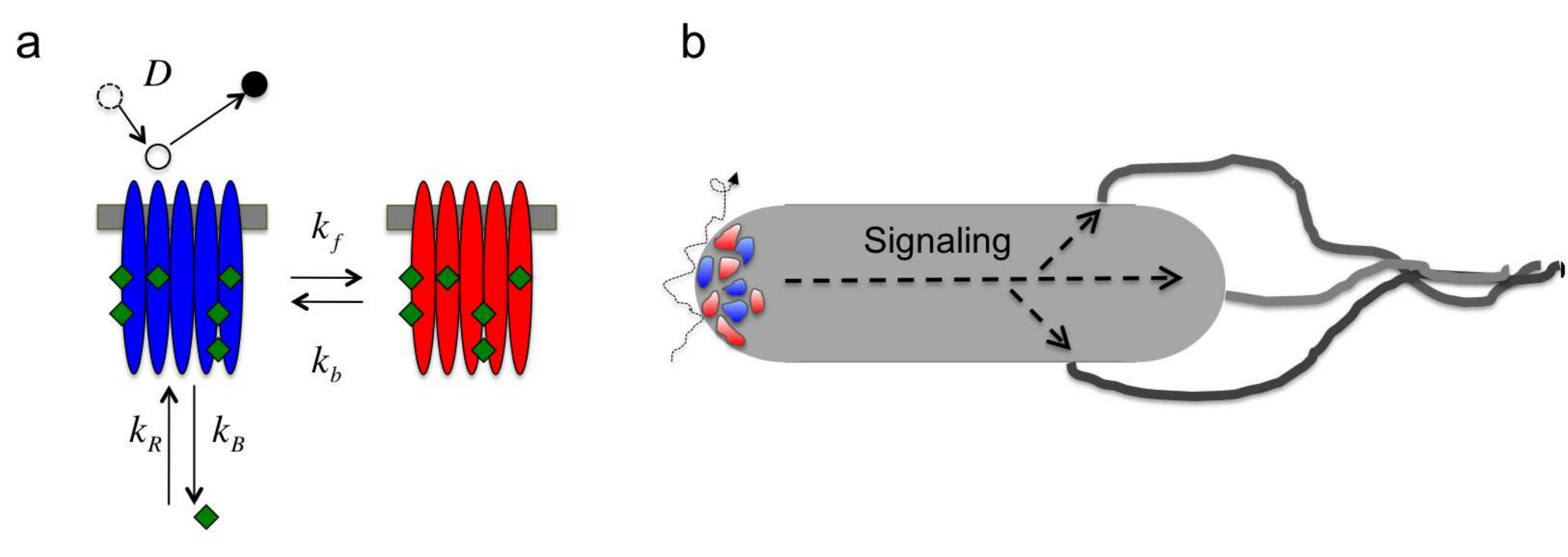}
\caption{\label{fig1} (color online) Schematic of bacterial cell with polar receptor cluster. (a) Receptor clusters are 
composed of smaller signaling complexes (here exemplified for $N=5$ receptors), which are either off/inactive (blue/left)
on/active (red/right). Ligand molecules arrive at receptors by diffusion (with diffusion constant $D$) and
bind/unbind. Receptor complexes randomly switch between the two states, from off to on 
with rate constant $k_f$ and from on to off with rate constant $k_b$. Additionally, receptors adapt by 
receptor methylation (rate constant $k_R$) and demethylation (rate constant $k_B$). Diamonds (green) indicate 
methyl-groups on receptors. (b) Shown is a trajectory of a diffusing ligand molecule, which comes on and 
off of the receptor cluster with $N_C=9$ complexes (and hence $N_T=N_C\cdot N=45$ receptors in total). 
The chemotaxis signaling pathway ultimately regulates the flagellated rotary motors for cell motility.}
\end{figure*}

\begin{figure}[t]
\includegraphics[width=8.5cm,angle=0]{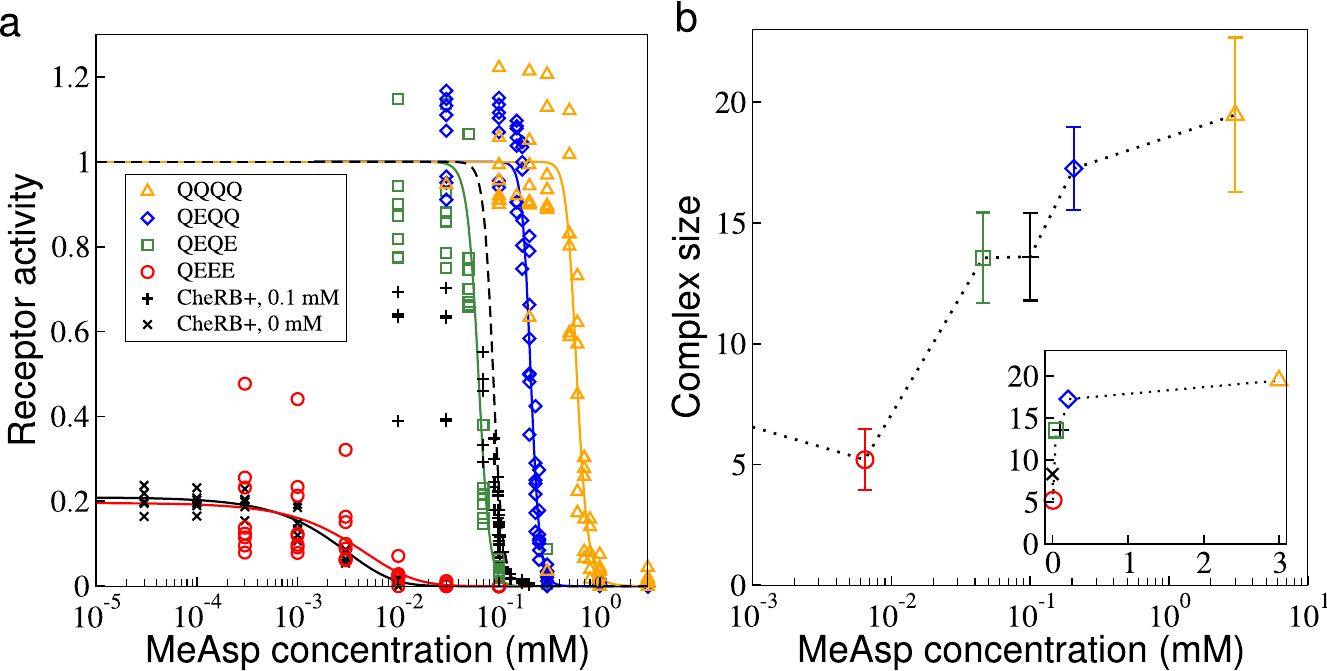}
\caption{\label{fig2} (color online) Data of chemotaxis signaling. (a) 
Dose-response curves as measured by {\it in vivo} FRET (symbols) and corresponding fits by the MWC model (solid 
lines) for {\it E. coli} cells expressing only Tar receptors. Cell types include adapting (CheRB+) and non-adapting, 
engineered cheRcheB mutants (QEEE, QEQE, QEQQ, and QQQQ with glutamate (E) or glutamine (Q) at four specific 
receptor modification sites). CheRB+ cells are adapted either to zero attractant (x symbols) or to 0.1 mM 
MeAsp (+ symbols). (b) Corresponding receptor complex sizes with 95.4\% confidence intervals, as extracted 
from the fitted MWC model. ({\it inset}) Same on linear plot to resolve zero ambient curve.  
Data and model curves, fitted with Principal Component Analysis, are reproduced from \cite{end08a}.}
\end{figure}

Receptor clustering is well documented in bacterial chemotaxis \cite{sou04b} and 
is known to amplify tiny changes in ligand concentration similar to an antenna.  
Experimental evidence for clustering is based on structural approaches \cite{ame02}, imaging by 
fluorescence microscopy \cite{thi07}, including photo-activated localization microscopy 
(PALM) \cite{gre09}, as well as cryo-electron microscopy \cite{bri08,khu08}. Receptor clusters 
form predominantly at the cell poles as illustrated in Fig.~\ref{fig1}, possibly due to the increased 
membrane curvature \cite{end09a}. At a smaller scale, receptor clusters are composed of smaller 
signaling complexes. The notion of receptor complexes is supported by high-resolution imaging with 
PALM \cite{gre09}, as well as by the extracted sensitivity and cooperativity from dose-response curves 
(activity changes in response to ligand stimuli) measured by {\it in vivo} 
fluorescence resonance energy transfer (FRET). As an example, Fig.~\ref{fig2}a shows previously 
published dose-response curves of the receptor activity from {\it in vivo} FRET experiments 
(see figure caption and \cite{end08a} for details). Briefly, cells were genetically 
engineered to only express the Tar receptor. Different curves correspond to different 
modification (adaptation) states of the receptors.

To explain the dose-response data, the Monod-Wyman-Changeux (MWC) model \cite{mon65} was used to successfully 
describe signaling by two-state receptor complexes (Fig.~\ref{fig1}a) \cite{sou04a,mel05,key06,end08a}. 
The complex size, {\it i.e.} the number of strongly coupled receptors in a complex, was estimated to 
be about 10-20 receptors. Alternative receptor models, 
later found to be inconsistent with the FRET data \cite{sko06}, are based on the Ising lattice, 
where moderate receptor-receptor coupling provides a mechanism for signal amplification and 
integration \cite{duk99,mel03}. Fits of the MWC model to the data are shown in Fig.~\ref{fig2}a, 
which indicates an increase in complex size with receptor methylation level and hence
ligand concentration (Fig.~\ref{fig2}b). This result is consistent with the observed destabilization 
of polar receptor clusters by receptor demethylation or addition of attractant \cite{end09a}. 
However, it is unknown what determines complex size.

Complex size could be limited by an imperfect physical clustering mechanism as proteins and lipids 
are soft materials, undergoing substantial thermal motion. Furthermore, larger complexes may not form 
due to the presence of other proteins in the membrane, which may constitute impurities in the 
receptor cluster. The dynamic aspect of receptors is supported by experiments using fluorescence 
recovery after photobleaching (FRAP). This indicates that receptor-cluster associated proteins, 
as well as components of the motors are relatively dynamic \cite{sch08,del09}. Alternatively, 
complex size might be determined by engineering principles (functionality), and hence be ``optimal'' 
for sensing. This work supports the latter view.

\begin{figure*}[t]
\includegraphics[width=15cm,angle=0]{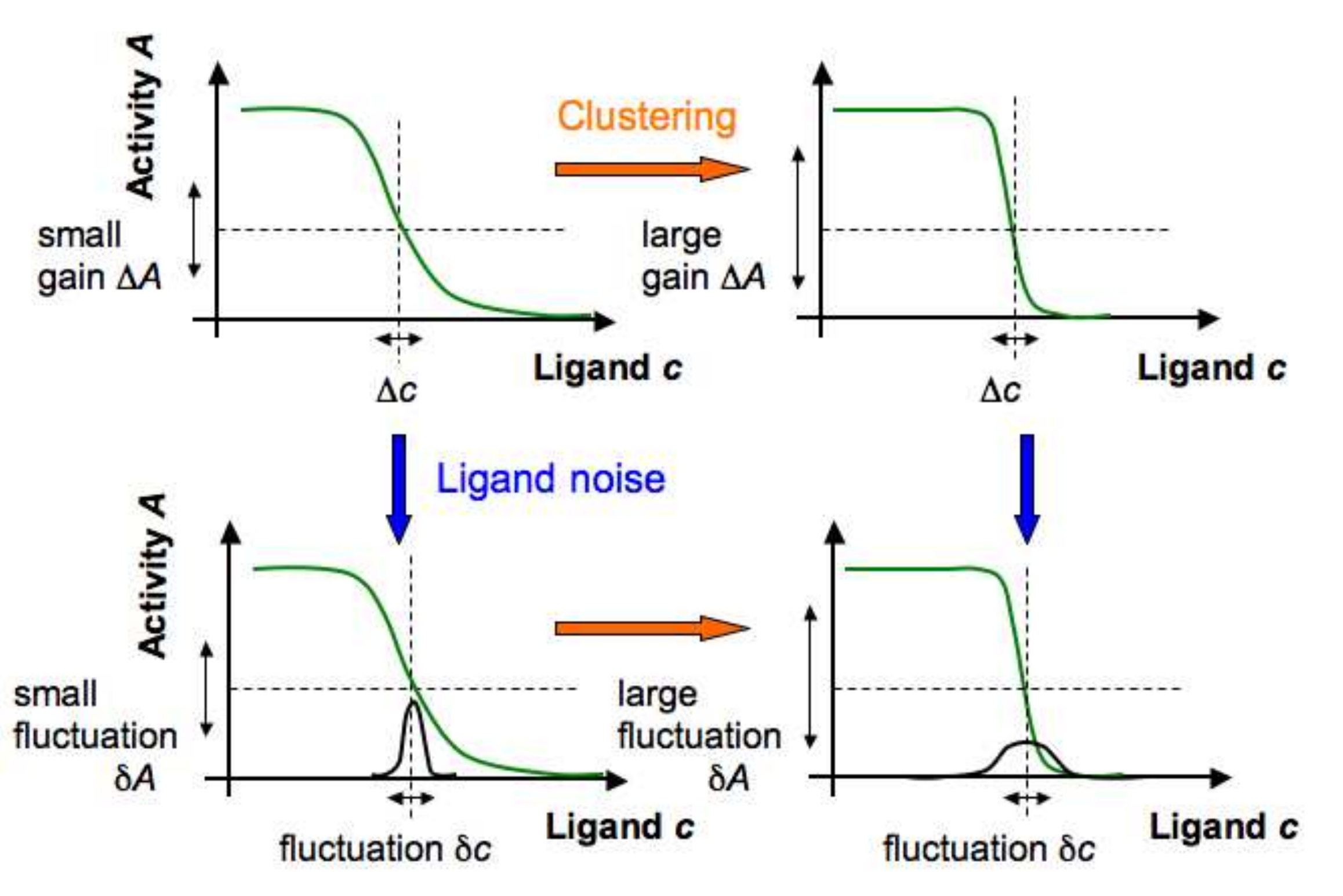}
\caption{\label{fig3} (color online) Schematic of dose-response curves of receptor activity. ({\it top panels}) 
Receptor cooperativity leads to steeper curves and hence larger amplification $\Delta A$ of a small stimulus, 
given by a change in ligand concentration $\Delta c$. ({\it bottom panels}) Noise in the stimulus, 
$\delta c$, represented by a peaked distribution, is amplified by receptor cooperativity as well, 
indicated by a fluctuation in activity $\delta A$. Receptor clustering, a potential side-effect of 
complex formation, leads to a further increase in input noise, shown by a broader distribution on 
the right.
}
\end{figure*}

Fig.~\ref{fig3} illustrates some of the advantages and disadvantages of receptor clustering. 
On the one hand, more receptor cooperativity, {\it i.e.} larger complexes, amplify signals better 
in the absence of input noise (top panels). On the other hand, random fluctuations 
in ligand concentration also become amplified by the complex 
(bottom panels). Furthermore, the closer the proximity between receptors in a cluster, the larger the 
fluctuations in ligand concentration for the cell, because nearby receptor complexes measure previously 
bound ligand molecules due to rebinding. Hence, clustering may render complexes highly prone to 
noise and reduce the cell's signal processing capabilities. Indeed, sources of noise are ubiquitous 
in biological sensing. 

Extrinsic input noise arises from the random arrival of ligand molecules at the cell-surface 
receptors, constituting the fundamental physical limit on concentration sensing, derived by Berg \& Purcell 
in 1977 \cite{ber77} and subsequently by others \cite{bia05,rap06,end08b,end09b,end09c}.
Specifically, Bialek \& Setayeshgar applied the Fluctuation-Dissipation Theorem (FDT) \cite{kub66} 
to derive the uncertainty in ligand sensing by receptors from the fluctuations in receptor occupancy.
Furthermore, if previously bound ligand molecules are removed, the uncertainty is significantly 
decreased \cite{end08b}. Such removal prevents ligand molecules from rebinding the receptors, and hence 
overcounting of the same ligand molecules by the cell. A potential mechanism for ligand removal is receptor
internalization, {\it e.g.} by endocytosis of ligand-bound receptors in eukaryotic cells \cite{aqu10}.

In addition to extrinsic noise, there is intrinsic noise in the signaling pathway, including the 
random receptor-complex switching between the {\it on} (active)  and {\it off} (inactive) 
states (similar to flickering of ion channels), 
as well as random receptor methylation and demethylation events \cite{Nis10}. Since the concentrations of methylation 
enzyme CheR and demethylation enzyme CheB are low, {\it i.e.} about a hundred copies per cell \cite{li04}, 
the fluctuations in receptor methylation level are expected to be significant \cite{kor04}.
To compensate for their small numbers, enzymes were found to transiently tether to the receptors.
This allows them to act on groups of 6-8 receptors \cite{li05}, reducing the noise in receptor 
methylation level due to the larger number of available modification sites \cite{end06}. In 
addition to these random biochemical events, there are further downstream signaling events, 
ultimately the random switching of the motors between its two rotational states.

How is signaling by receptor complexes affected by extrinsic and intrinsic noise? In this work, we 
use the well-characterized example of bacterial chemotaxis to combine the allosteric MWC model 
for signaling by receptor complexes \cite{key06,end08a} with calculations of noise to study their 
interconnectedness. Using the FDT \cite{kub66,bia05}, we calculate the uncertainty in
ligand concentration sensing by the cell. Specifically, we consider the effects of the random arrival of 
ligand molecules at the receptors by diffusion and rebinding, switching of the receptor complexes, 
and receptor methylation/demethylation. While these effects have been described individually before, 
we combine these to address signaling by multiple receptor 
complexes in a cell. Based on a simplified model, we then calculate the signal-to-noise ratio (SNR), 
summarizing the balance of beneficial and detrimental effects of clustering for the cell. Interestingly, 
we find that there is no advantage for the cell to assemble receptor complexes for noisy input stimuli 
in the absence of intrinsic signaling noise. However, with such intrinsic noise included, an optimal complex 
size arises in line with estimates of the sizes of chemoreceptor complexes in bacteria and protein 
aggregates in eukaryotic cells.

The paper is organized as follows: In Section \ref{sec2} we describe amplification of stimuli and
extrinsic noise by receptor complexes. In section \ref{sec3}, we derive the uncertainty in 
ligand concentration sensing by multiple receptor complexes in the cell. In section \ref{sec4} 
we combine the information provided in sections \ref{sec2} and \ref{sec3} to derive an optimal complex 
size, determined by the balance between signal and noise amplification by the receptor complexes. We 
conclude with final comments and discussion in section \ref{sec5}. Furthermore, appendix \ref{secA}
provides details on our model and, starting from the Master equation, derives the noise terms 
using the van Kampen expansion. Appendix \ref{secB} is devoted to summarizing the parameter values 
used. In appendix \ref{secC} we examine the effect of receptor distribution on the uncertainty of 
sensing.

\section{\label{sec2} Stimulus and noise amplification}

Signaling in bacterial chemotaxis is quantitatively interpreted within the MWC model. 
In this model, receptors form signaling complexes, believed to consist of about 10-20 receptors. 
Due to strong receptor-receptor coupling within a complex, a complex is an effective two-state 
system with all receptors either on or off together.
Specifically, we consider MeAsp-binding to complexes of the Tar receptor in line with
recent experiments \cite{end08a}. Since MeAsp binds more favorably to the receptor off than 
to the receptor on state, ligand generally tends to turn the receptor activity off, whereas receptor 
methylation favors the on state and so compensates for ligand binding during adaptation. 

In the MWC model, the probability that a receptor complex is active, {\it i.e.} the receptor activity, 
$A$, depends only on the free-energy difference $F$ (free energy from now on) between its on and 
off states \cite{key06,end08a}
\begin{equation}
A=\frac{1}{1+e^{F}}\label{Pon}
\end{equation}
with energies in units of thermal energy $k_BT$. In this model, for a complex size of
$N$ receptors, the complex free energy is simply $N$ times the free energy of a single receptor
\begin{equation}
 F=N\left[E+\ln\left(\frac{1+c/K_D^\text{off}}{1+c/K_D^\text{on}}\right)\right]\label{Fm}
\end{equation}
with ligand concentration $c$ and ligand dissociation constants $K_D^\text{on}$ and $K_D^\text{off}$ 
for the on and off states, respectively. These constants represent the ligand concentrations at 
which the receptor in each state is occupied by ligand with 50 percent probability.
In the absence of ligand, the free energy of a receptor is given by $E=\alpha-\beta\,m_1$ 
with $m_1$ the methylation level (methyl-group concentration) corresponding to a single receptor, 
and parameters $\alpha$ and $\beta$ recently determined
for the Tar receptor \cite{end08a}. Eq.~\eqref{Fm} can be written in terms of the total methylation 
level (concentration) of the whole complex using $m=Nm_1$, resulting in
\begin{equation}
 F=N\alpha-\beta m\, +\, N\ln\left(\frac{1+c/K_D^\text{off}}{1+c/K_D^\text{on}}\right).\label{Fm2}
\end{equation}
This model has been very successful in describing stimulus amplification, 
precise adaptation to persistent stimulation, and signal integration by mixed receptor types. 
For instance, the MWC model is able to describe the dose-response curves in Fig.~\ref{fig2} 
\cite{end08a} and other data \cite{key06,sko06,end07}.

In cells adapted to average steady-state activity $\bar A$, the methylation level $\bar m$ is 
determined by precise adaptation to ligand concentration $\bar c$ via
\begin{equation}
\bar m=\frac{1}{\beta}\left[N\alpha -\ln(1/\bar A-1)
-N\ln\left(\frac{1+\bar c/K_D^\text{on}}{1+\bar c/K_D^\text{off}}\right)\right].
\end{equation}
The mechanism for the cell to achieve precise adaptation
was originally proposed by Barkai \& Leibler \cite{bar97} and was later identified as 
{\it integral feedback control} \cite{yi00}.
Briefly, if the dynamics of the methylation level are independent of the available modification sites
and external ligand concentration, then the adapted steady-state activity only depends on cell-specific 
parameters. Specifically, the dynamics of adaptation were recently determined \cite{cla10}
\begin{equation}
 \frac{dm}{dt} = k_R (1-A) - k_B A^3,\label{dMdt}
\end{equation}
where $k_R$ ($k_B$) is the rate constant of methylation (demethylation) by enzymes 
CheR (CheB). Eq.~\eqref{dMdt} assumes that CheR only methylates active receptors and CheB only 
demethylates inactive receptors in line with experimental observation.
Furthermore, for demethylation CheB needs to be activated by phosphorylation and may act cooperatively with
other CheB enzymes, explaining the $A^3$ dependence in Eq.~\eqref{dMdt} \cite{cla10}.

\begin{figure}[t]
\includegraphics[width=8.5cm,angle=0]{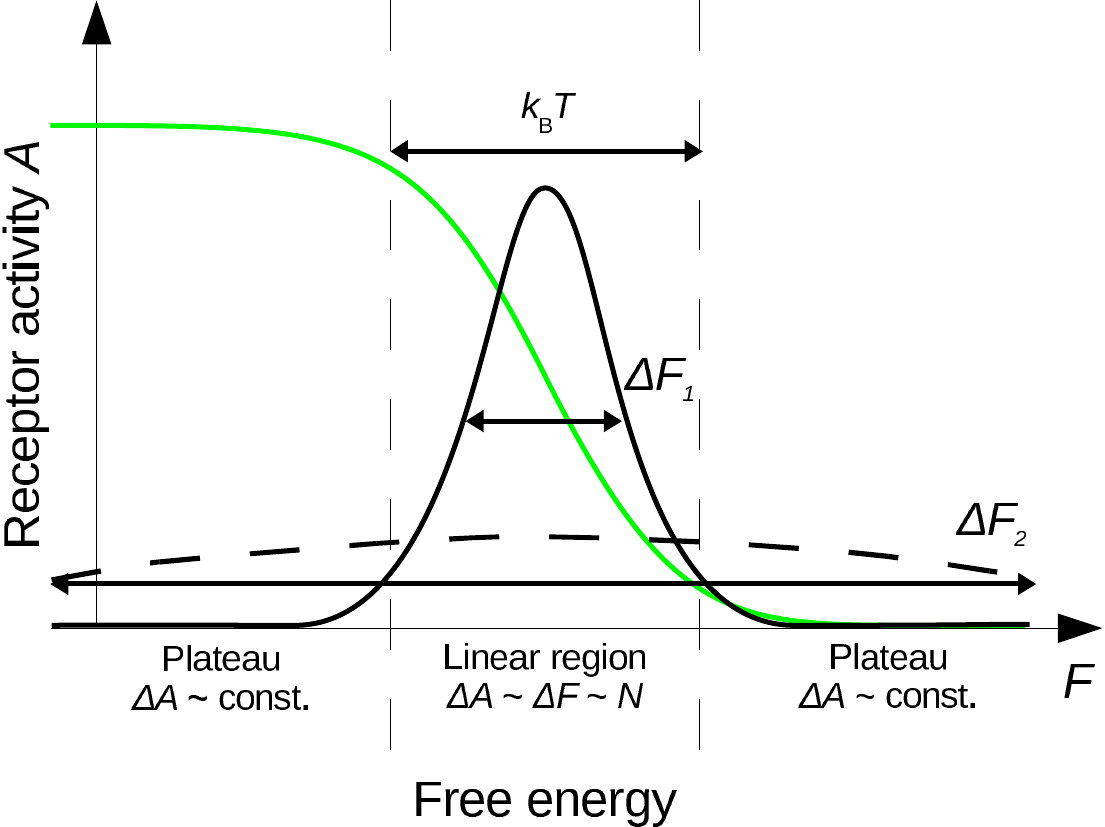}
\caption{\label{fig4} (color online) Activity of receptor complex $A(F)=1/(1+\exp(F))$ in green (light gray) 
and distribution of signaling complexes as a function of complex free energy $F$ (solid black line: 
small changes $\Delta F_1$, dashed black line: large changes $\Delta F_2$). Free energies in units of 
thermal energy $k_BT$, indicated by top arrow.}
\end{figure}

For initially adapted cells, signal amplification is obtained by expanding the activity in terms of 
a small stimulus $\Delta c$. In the linear regime, the change in receptor-complex activity is given by
\begin{equation}
\Delta A=\left(\frac{\partial A}{\partial F}\right)\left(\frac{\partial F}{\partial c}\right)\Delta c
=-N\bar A(1-\bar A)\Delta n\frac{\Delta c}{\bar c}\label{DA}
\end{equation}
with $\partial A/\partial F=-\bar A(1-\bar A)$, $\partial F/\partial c=
N\Delta n/\bar c$, and
\begin{equation}
\Delta n=\frac{\bar c}{\bar c+K_D^\text{off}}-\frac{\bar c}
{\bar c+K_D^\text{on}}\label{Dn}
\end{equation}
the difference in receptor occupancy between its on and off states. Hence, due to receptor cooperativity 
in the MWC model the response $\Delta A$ corresponds to an amplification of small stimuli by complex 
size $N$. For larger stimuli, 
the response $\Delta A$ saturates to zero or maximal activity (Fig.~\ref{fig3}, top panels), which occurs when 
the associated free-energy change is comparable to the thermal energy ($\Delta F\approx 1$, see
Fig.~\ref{fig4}). As a consequence, a proper investigation of signaling by receptor complexes requires the 
full non-linear expression for the activity in Eq.~\eqref{Pon}.

Importantly, Eq.~\eqref{DA} also applies to amplification of ligand noise, {\it i.e.} $\delta A\propto N\delta c$
with $\delta c$ describing a small fluctuation in ligand concentration, indicating that 
receptor complex formation and cooperativity also have a detrimental effect (Fig.~\ref{fig3}, 
bottom panels). 

\section{\label{sec3} Uncertainty of sensing}
 
The uncertainty in sensing ligand concentration stems from extrinsic noise (random arrival of
ligand molecules at receptors by diffusion and their rebinding), as well as intrinsic 
signaling noise (receptor complex switching and methylation/demethylation), projected 
outside the cell in form of extrinsic noise in disguise. If we 
neglect cross correlations (in this section for illustrative purposes), the uncertainty
has the form $\langle(\delta c)^2\rangle_\tau=\sum_k\langle(\delta c)^2\rangle_\tau^\text{k}$
with contributions $k$ and $\tau$ an averaging time of the noise due to 
slower downstream signaling. The aim of this section is to demonstrate how the different
contributions are affected by receptor complex size. Specifically, we show which contributions 
are amplified and which ones are not. This will help classifying noise sources more effectively, 
and guide the investigation of the optimal complex size in the next section.

Bialek \& Setayeshgar recently calculated the uncertainty in ligand concentration using a
single MWC complex as a biological measurement device \cite{bia08}. In their model, the slow 
(compared to ligand binding and unbinding) random switching between the on and off states of 
the complex leads to a release or uptake of several ($N\Delta n$) ligand molecules, since both 
states are characterized by different ligand dissociation constants. Here, we first extend the 
model by Bialek \& Setayeshgar \cite{bia08} to multiple MWC complexes of Tar receptors.
Subsequently, we apply the method to intrinsic methylation noise, and also 
discuss the other noise contributions to the uncertainty.

The dependence of the activity on the switching of the receptor complex is described by
\begin{equation}
\frac{dA(t)}{dt}=k_f[1-A(t)]-k_bA(t)\label{dAdt1}
\end{equation}
with the forward and backward rate constants given by $k_f$ and $k_b$, respectively ({\it cf.} 
Fig.~\ref{fig1}a). The resulting steady-state value for the activity is given by
\begin{equation}
\bar A=\frac{k_f}{k_f+k_b}=\frac{1}{1+e^{\bar F}}
\end{equation}
with 
\begin{equation}
e^{\bar F}=\frac{k_b}{k_f}.\label{kF}
\end{equation}
Now, we consider $N_C$ such receptor complexes, which, due to a change in receptor occupancy from 
switching, couple to the ligand diffusion equation
\begin{subequations}
\begin{align}
\!\!\!\!\!\frac{dA_{j}(t)}{dt}&=k_f[1-A_{j}(t)]-k_bA_{j}(t)\label{dAdt}\\
\!\!\!\!\!\frac{\partial c(\vec x,t)}{\partial t}&=D\nabla^2c(\vec x,t)+N\Delta n
\sum_{l=1}^{N_C}\delta(\vec x-\vec x_l)\frac{dA_{l}(t)}{dt}\label{dcdt},
\end{align}
\end{subequations}
where $A_{j}$ is the receptor-complex activity of the $j$th complex at position $\vec x_j$. Furthermore,
in Eq.~\eqref{dcdt} parameter $D$ is the diffusion constant and $\delta(\dots)$ is the Dirac delta
function to describe the location of the complexes.

Following Refs. \cite{bia05,end09b}, we linearize Eqs.~\eqref{dAdt} and \eqref{dcdt}
around the steady-state receptor-complex activity and ligand concentration via
$\delta A=A(t)-\bar A$ and $\delta c(t)=c(t)-\bar c$, respectively. We further linearize the
rate constants $\delta k_f(t)=k_f(t)-\bar k_f$ and $\delta k_b(t)=k_b(t)-\bar k_b$,
allowing us to apply the FDT below \cite{bia05}. Specifically, we replace 
the fluctuations in rate constants by fluctuations in their conjugate variable, {\it i.e.} the
receptor-complex free energy \cite{bia05}, by using
\begin{subequations}
\begin{align}
\delta F&=\delta E+N\Delta n\frac{\delta c}{\bar c}\nonumber\\
&=\frac{\delta k_b}{\bar k_b}-\frac{\delta k_f}{\bar k_f},\label{dF}
\end{align}
\end{subequations}
obtained by linearizing Eqs.~\eqref{Fm2} and \eqref{kF}.

Next, we Fourier Transform the linearized equations $d(\delta A_{j})/dt$ and $\partial(\delta c)/\partial t$
into frequency and wave-vector space, defined by $G(\vec x, t)=
\int\frac{d\omega}{2\pi}\frac{dk^3}{(2\pi)^3}\exp\{i(\vec k\vec x-\omega t)\}\hat G(\omega,\vec k)$ for any
integrable function $G$. This results in an equation for frequency-dependent fluctuations in the activity of 
the $j$-th complex
\begin{eqnarray}
&&\!\!\!\!\!\!\!\!\!\!\!(\bar k_f+\bar k_b-i\omega)\delta\hat A_{j}(\omega)\nonumber\\
&&\quad\quad=-\bar k_b\bar A_{j}\left[\delta\hat E_j
+N\Delta n\frac{\delta\hat c(\vec x_j,\omega)}{\bar c}\right],\label{dhatA}
\end{eqnarray}
and the wavevector and fequency-dependent variation in ligand concentration
\begin{equation}
\delta\hat c(\vec k,\omega)=\frac{-i\omega N\Delta n}{(Dk^2-i\omega)}\sum_{l=1}^{N_C}e^{-i\vec k\vec x_l}\delta\hat A_{l}.\label{dhatc}
\end{equation}
To remove the $\vec x$ dependence in Eq.~\eqref{dhatA}, we invert the spatial Fourier Transform in Eq.~\eqref{dhatc}, 
resulting in
\begin{equation}
\delta\hat c(\vec x_j,\omega)=\frac{-i\omega N\Delta n}{\pi^2D}\left[\frac{1}{2a}\delta\hat A_{j}(\omega)+
\frac{\pi}{4}\sum_{l\neq j}^{N_C}\frac{\delta\hat A_{l}(\omega)}{|\vec x_j-\vec x_l|}\right],\label{dcx}
\end{equation}
where the receptor complex dimension $a$ was introduced to regularize an integral. Eq.~\eqref{dcx}
is valid for low frequencies $\omega<\!\!<D/a^2$, {\it i.e.} we assume the time to read out the receptor free energy to be
long compared to the correlation time between receptor-complex switching events \cite{bia08}. 
Inserting this equation into Eq.~\eqref{dhatA}, we obtain
\begin{eqnarray}
\left[\bar k_f+\bar k_b-i\omega\left(1+\bar k_b\bar A\frac{N^2\Delta n^2}
{2\pi Da\bar c}\right)\right]\delta\hat A_{j}(\omega)\nonumber\\
=i\omega\bar k_b \bar A\frac{N^2\Delta n^2}{4\pi D\bar c}\sum_{l\neq j}^{N_C}
\frac{\delta\hat A_{l}(\omega)}{|\vec x_j-\vec x_l|}-\bar k_b\bar A\delta\hat E_j(\omega).
\end{eqnarray}
Next, we sum over all receptor complexes using $N_C\delta\hat A=\sum_{j=1}^{N_C}\delta\hat A_{j}$ and
$\delta\hat E=\sum_{j=1}^{N_C}\delta\hat E_j$ to obtain the total receptor activity and free energy. Furthermore,
we introduce the geometric structure factor $\Phi=\sum_{j\neq1}^{N_C}\frac{1}{|\vec x_1-\vec x_j|}$,
valid for receptor complex distributions for which each receptor complex is equivalent to all the other receptor
complexes (ring or sphere of receptors) \cite{bia05}. With these quantities introduced, we obtain
\begin{eqnarray}
&&\!\!\!\!\!\!\!\!\!N_C\left\{\bar k_f+\bar k_b-i\omega\left[
1+\frac{\bar k_b\bar A N^2\Delta n^2}{2\pi D\bar c}\left(\frac{1}{a}+\frac{\Phi}{2}\right)
\right]\right\}\delta\hat A\nonumber\\
&&\quad\quad=-\bar k_b\bar A\delta\hat E.
\end{eqnarray}
Using the FDT, we calculate the noise power spectrum of the receptor-complex
activity, defined by $\langle\delta\hat A(\omega)\delta\hat A^*(\omega)\rangle$, 
from the deterministic linear response to a small perturbation in the receptor-complex free energy
\begin{subequations}
\begin{align}
S_{A}(\omega)=&\frac{2}{\omega}\text{Im}\left[-\frac{\delta\hat A}{\delta\hat E}\right]\label{w}\\
=&\frac{2\bar k_f(1-\bar A)(1+\Sigma)}{N_C[(\bar k_f+\bar k_b)^2+
\omega^2(1+\Sigma)^2]}\\
\stackrel{\omega\rightarrow0}{\longrightarrow}&\frac{2\bar k_f(1-\bar A)(1+\Sigma)}
{N_C(\bar k_f+\bar k_b)^2}\label{w0}
\end{align}
\end{subequations}
with $\Sigma=\frac{\bar k_b\bar A N^2\Delta n^2}{2\pi D\bar c}(1/a+\Phi/2)$ 
and Eq.~\eqref{w0} valid in the zero-frequency limit. Note the minus sign in Eq.~\eqref{w} is 
introduced since a positive $\delta \hat E$ leads to a negative $\delta\hat A$ \cite{bia08}. 
From $\delta A=-N\bar A(1-\bar A)\Delta n\frac{\delta c}{\bar c}$ 
({\it cf.} Eq.~\eqref{DA}), we obtain for the time-averaged variance of the ligand concentration
\begin{equation}
\langle(\delta c)^2\rangle_\tau=\left[\frac{\bar c}{N\Delta n\bar A(1-\bar A)}\right]^2
\frac{S_A(0)}{\tau}
\end{equation}
with $\tau$ the averaging time determined by slow, downstream signaling, and finally 
for the relative uncertainty in sensing
\begin{eqnarray}
\frac{\langle(\delta c)^2\rangle_\tau^\text{SR}}{\bar c^2}&=&\frac{2}{N_CN^2\Delta n^2\bar k_f
(1-\bar A)\tau}\nonumber\\
&&\quad+\frac{1}{N_C\pi D\bar c\tau}\left(\frac{1}{a}+\frac{\Phi}{2}\right).\label{dcSR}
\end{eqnarray}
The first term on the right-hand side represents the uncertainty in ligand concentration due to
the release and uptake of $N\Delta n$ ligand molecules induced by the randomly switching receptor
complexes (S). The second term is due to diffusion and represents the additional uncertainly from rebinding 
of previously bound ligand molecules (R). 
Due to this term, the uncertainty depends on the spatial distribution of the receptor complexes on the 
cell surface. Specifically, the term proportional to $1/a$ describes the rebinding of ligand molecules to the 
same receptor complex, while the term proportional to $\Phi$ describes the rebinding to the other receptor 
complexes. This latter contribution becomes the larger the smaller the proximity of the receptor complexes, 
{\it e.g}. in the polar receptor cluster (appendix \ref{secC}). Fast ligand diffusion (or removal of 
bound ligand molecules by an efficient cellular uptake mechanism) reduces this term \cite{aqu10}. 
Furthermore, in Eq.~\eqref{dcSR} the number of receptor complexes $N_C$ in the denominators reduces the 
uncertainty by spatial averaging.

The FDT method can also be used to calculate the uncertainty in ligand concentration from random receptor 
methylation and demethylation events. The rate of change of a small deviation of the receptor-complex 
activity $\delta A$ due to a change in total receptor methylation level $\delta m$ is given by
\begin{equation}
\frac{d(\delta A)}{dt}=\frac{\partial A}{\partial m}\frac{d(\delta m)}{dt}\label{dAdt2}
\end{equation}
where $dm/dt$ and $d(\delta m)/dt$ are given by Eq.~\eqref{dMdt} and its linearised version, 
respectively. Using
\begin{eqnarray}
\delta F&=&\frac{\delta k_R}{\bar A(\bar k_R+3\bar k_B\bar A^2)}\nonumber\\
&&\quad+\frac{\bar A^2}{1-\bar A}\cdot\frac{\delta k_B}{\bar k_R+3\bar k_B\bar A^2},
\end{eqnarray}
linearization and Fourier transformation finally leads to the relative uncertainty in ligand 
concentration
\begin{eqnarray}
\frac{\langle(\delta c)^2\rangle_\tau^\text{MR}}{\bar c^2}&=&\frac{2}{N_CU^2\Delta n^2\bar k_R
(1-\bar A)\tau}\nonumber\\
&&\quad+\frac{1}{N_C\pi D\bar c\tau}\left(\frac{1}{a}+\frac{\Phi}{2}\right),\label{dcMD}
\end{eqnarray}
where the first term on the right hand side describes the contribution from random receptor
methylation and demethylation events (M), respectively leading to a release and take-up of 
ligand molecules. This term is inversely proportial to $N^2$, and hence, is not
amplified by receptor cooperativity (cf. Eq. (6))  in analogy to the intrinsic ligand
noise arising from random switching of the receptor complex in Eq.~\eqref{dcSR}.
The second term in Eq.~\eqref{dcMD} is idential to Eq.~\eqref{dcSR} and describes the contribution 
from diffusion (R), as released and taken up ligand molecules lead to additional uncertainty in
ligand concentration.

So far the contribution to the uncertainty from random binding and unbinding of ligand molecules (L) 
to the receptor complex is still missing. To avoid the complexity of different rates for the on and off states, 
we assume diffusion-limited binding to the receptor cluster and write for the additional 
uncertainty \cite{end09b,aqu10}
\begin{equation}
\frac{\langle(\delta c)^2\rangle_\tau^\text{L}}{\bar c^2}=
\frac{1}{4\pi DR_s\bar c\tau}\label{dcL},
\end{equation}
calculated from Poisson statistics and the diffusive flux to an absorbing sphere of radius $R_s$, which 
represents the dimension of the receptor cluster. 
Eq.~\eqref{dcL} is considered the fundamental physical limit of sensing as it cannot be reduced 
by any intracellular sensing mechanism. This extrinsic noise is amplified due to the absence of a $N^2$ factor in the denominator. 

In summary, intrinsic and extrinsic noise affect the uncertainty of sensing differently, 
{\it i.e.} only extrinsic noise is amplified.
In the next section, we consider an integrative model of extrinsic ligand noise and intrinsic noise from 
receptor methylation/demethylation noise. The receptor-complex 
switching noise is much smaller due to the large switching rates and 
hence is assumed to be averaged out. By using the receptor activity of the whole cell as a 
read-out of signaling, we are able to compare the properties of stimulus and noise transmission.\\

\section{\label{sec4} Optimal receptor complex size}

What effects have stimulus and noise amplification on the signaling capabilities of the whole cell, 
and specifically, is there an optimal complex size? 
In the cell, we assume a large receptor cluster of $N_\text{T}$ identical receptors 
divided into $N_C$ smaller receptor signaling complexes of $N$ Tar receptors each (see Fig.~\ref{fig1}b). 
We now calculate the $N$-dependent SNR for the total activity $A_T$ of the cell in response to 
a uniform, non-saturating stimulus $\Delta c\propto \bar c$
\begin{equation}
\text{SNR}=\frac{\text{Signal}}{\text{Noise}}=\frac{<\Delta A_\text{T}>_N^2}{<\!\!<\delta A_\text{T}^2>\!\!>_N},
\end{equation}
where the Signal is defined by the squared-mean response of all receptors in the cell
$<\!\!\Delta A_\text{T}\!\!>_N^2=\langle\sum_i^{N_T}\Delta A_i\rangle^2=(N_T\langle\Delta A\rangle)^2$
to the stimulus, neglecting cross-correlations between the activities of different complexes.
In contrast, the Noise is expressed by the mean-square deviation of the independently fluctuating 
receptor complexes $<\!\!<\delta A_\text{T}^2>\!\!>_N=\sum_j^{N_C}\langle\delta A_j^2\rangle=
N_C\langle\delta A^2\rangle$. Since measurements are not done instantaneously by the cell, we use
time-averaged activities $\langle...\rangle=\tau^{-1}\int_t^{t+\tau}....d\tilde t$. This leads to the
following general expressions for the Signal and the Noise
\begin{widetext}
\begin{eqnarray}
<\Delta A_\text{T}>_N^2&=&\left\{\frac{N_\text{T}}{\tau}\int_t^{t+\tau}
\left[ A(F(c(\tilde t)+\Delta c,m(\tilde t))-A(F(c(\tilde t),m(\tilde t))\right]\, d\tilde t\right\}^2\label{SC}
\end{eqnarray}

\begin{eqnarray}
<\!\!<\delta A_\text{T}^2>\!\!>_N&=&\frac{N_\text{C}}{\tau}\int_t^{t+\tau}
[N\,A(F(c(\tilde t),m(\tilde t)))-N\,\bar A]^2\, d\tilde t
=\frac{N_\text{T}N}{\tau}\int_t^{t+\tau}[A(F(c(\tilde t),m(\tilde t)))-\bar A]^2\, d\tilde t\label{NC}
\end{eqnarray}
\end{widetext}
with $A=A(F)$ the receptor complex activity, depending on ligand concentration $c$ and methylation level 
$m$ via the free energy $F=F(c,m)$. Both ligand concentration $c=c(\tilde t)$ and methylation level 
$m=m(\tilde t)$ depend on time. Further note that the factor $N$ on the right-hand side 
of Eq.~\eqref{NC} appears as each receptor in a complex has the same activity. 
Even though the stimulus and noise (included below) are small, Eqs.~\eqref{SC} and \eqref{NC}
allow for non-linear effects in the activity for the large complexes sizes considered.

For computational feasibility, we exploit the ergodic hypothesis, allowing us to replace the time averages 
by the ensemble averages. This leads to the respective Signal and Noise
\begin{widetext}
\begin{eqnarray}
 <\Delta A_\text{T}>_N^2&=&\left\{N_\text{T}\int \left[ A(F(c+\Delta c,m))-A(F(c,m))\right]
 \,P(c,m)\,dc\,dm\right\}^2\label{Num2}\\
 <\!\!<\delta A_\text{T}^2>\!\!>_N&=&N_\text{T}N\int[\,A(F(c,m))-\,\bar A]^2\,P(c,m)\,dc\,dm.\label{Denom2}
\end{eqnarray}
\end{widetext}
In the Signal, receptor complexes experience a stimulus $\Delta c$ on top of a fluctuating ligand concentration,
while in the Noise, fluctuations in activity are measured relative to the average activity. In Eqs. (\ref{Num2})
and (\ref{Denom2}), we further use a bivariate Normal distribution to describe the joint probability of the 
ligand concentration and methylation level at a receptor complex, given by
\begin{equation}
P(c,m)=\frac{e^{-\frac{1}{2(1-\rho^2)}
\left[\frac{(c-\bar c)^2}{\langle(\delta c)^2\rangle}+\frac{(m-\bar m)^2}{\langle(\delta m)^2\rangle}
-\frac{2\rho (c-\bar c)(m-\bar m)}{\sqrt{\langle(\delta c)^2\rangle}\sqrt{\langle(\delta m)^2\rangle}}\right]}}
{2\pi\sqrt{\langle(\delta c)^2\rangle}\sqrt{\langle(\delta m)^2\rangle}\sqrt{1-\rho^2}}.
\label{PcM}
\end{equation}
In addition to the variances $\langle(\delta c)^2\rangle$ and $\langle(\delta m)^2\rangle$, 
Eq.~\eqref{PcM} also depends on covariance $\langle(\delta c)(\delta m)\rangle$, included in the correlation coefficient
$\rho=\langle(\delta c)(\delta m)\rangle/[\sqrt{\langle(\delta c)^2\rangle}\sqrt{\langle(\delta m)^2\rangle}]$.
While the methylation level can fluctuate due to random methylation and demethylation events
independent from fluctuations in ligand concentration, fluctuations in ligand concentration can 
induce fluctuations in the methylation level due to adaptation (via parameter $\rho$).

To calculate the variances and covariance we use a simplified Master equation, describing how the receptor-complex 
activity depends on external ligand concentration and receptor methylation level. The ligand noise includes effects of
the random arrival of ligand molecules at the receptors and their rebinding by diffusion, given by rate $k_d=D/a^2$.
We then apply the van Kampen expansion to obtain the second moments of the joint distribution (see appendix \ref{secA} 
for details).

\subsection{Ligand noise}

First, we only consider extrinsic ligand noise by setting the methylation level $m(\tilde t)$ equal to the constant 
adapted value $\bar m$. The distribution of the ligand concentration is now effectively given by 
$P(c)=1/\sqrt{2\pi\langle(\delta c)^2\rangle}\exp\{-(c-\bar c)^2/[2\langle(\delta c)^2\rangle]\}$, 
assumed to be Normal with average ligand concentration $\bar c$ and variance
\begin{equation}
\langle(\delta c)^2\rangle=\frac{\bar c}{a^3}
\end{equation}
(see appendix \ref{secA} for more details). 

Importantly, the Signal and Noise have characteristic $N$-dependencies, which need to be examined in order
to answer the question of the optimal complex size. We first consider the linear regime of the 
activity, as this can be solved analytically. In this case, Eqs. (\ref{Num2}) and (\ref{Denom2}) reduce to
\begin{equation}
<\!\Delta A_\text{T}\!>_N^2=N_\text{T}^2\left(\frac{\partial A}{\partial c}\right)^2\Delta c^2\propto N^2,\label{Num11}
\end{equation}
and
\begin{equation}
<\!\!<\delta A_\text{T}^2>\!\!>_N=N_\text{T}N\left(\frac{\partial A}{\partial c}\right)^2\langle(\delta c)^2\rangle\propto N^3,\label{Denom11}
\end{equation}
using Eq.~\eqref{DA}. As a result, the SNR scales as $N^{-1}$ and hence decreases for increasing complex size.
This indicates that a single receptor is better than a complex of multiple receptors for signaling
due to the more rapid increase of the Noise with complex size than the Signal. Note also that the SNR
is proportional to the total number of receptors $N_T$ in a cell. 

\begin{figure}[t]
\includegraphics[width=9cm,angle=0]{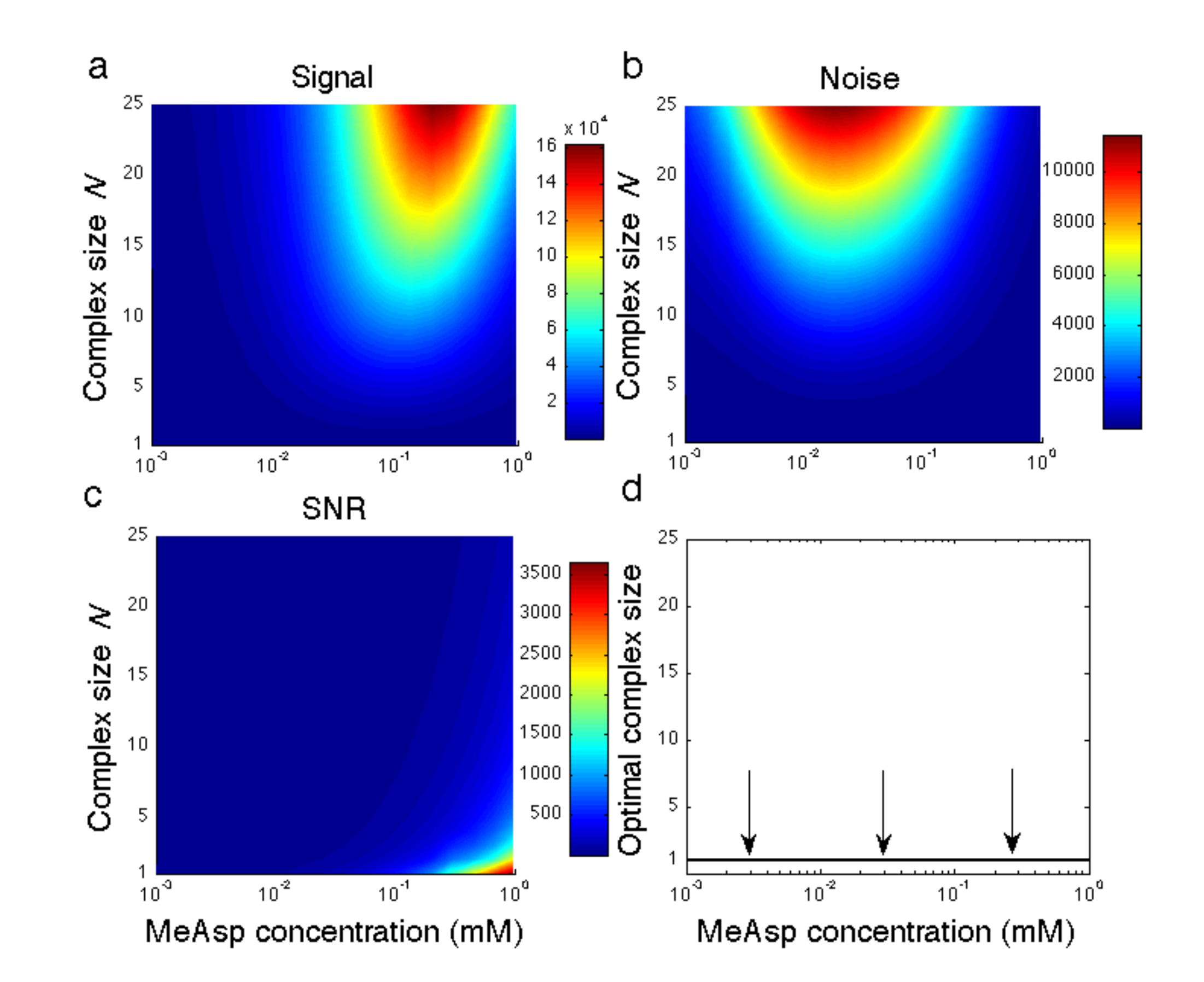}
\caption{\label{fig5} 
(color online) Absence of an optimal complex size for extrinsic ligand noise. (a) Signal, (b) Noise, and (c) SNR 
as a function of MeAsp concentration and complex size, {\it i.e.} the number of Tar receptors 
in a complex. (d) There is no maximum of the SNR for complex sizes larger than one receptor. 
For each MeAsp concentration, the same non-saturating MeAsp stimulus is applied to all complex sizes.
Parameter values are given in appendix \ref{secB}. Integration performed with the 
quadrature method in Matlab (Mathworks, Natwick, MA). }
\end{figure}

\begin{figure}[t]
\includegraphics[width=9cm,angle=0]{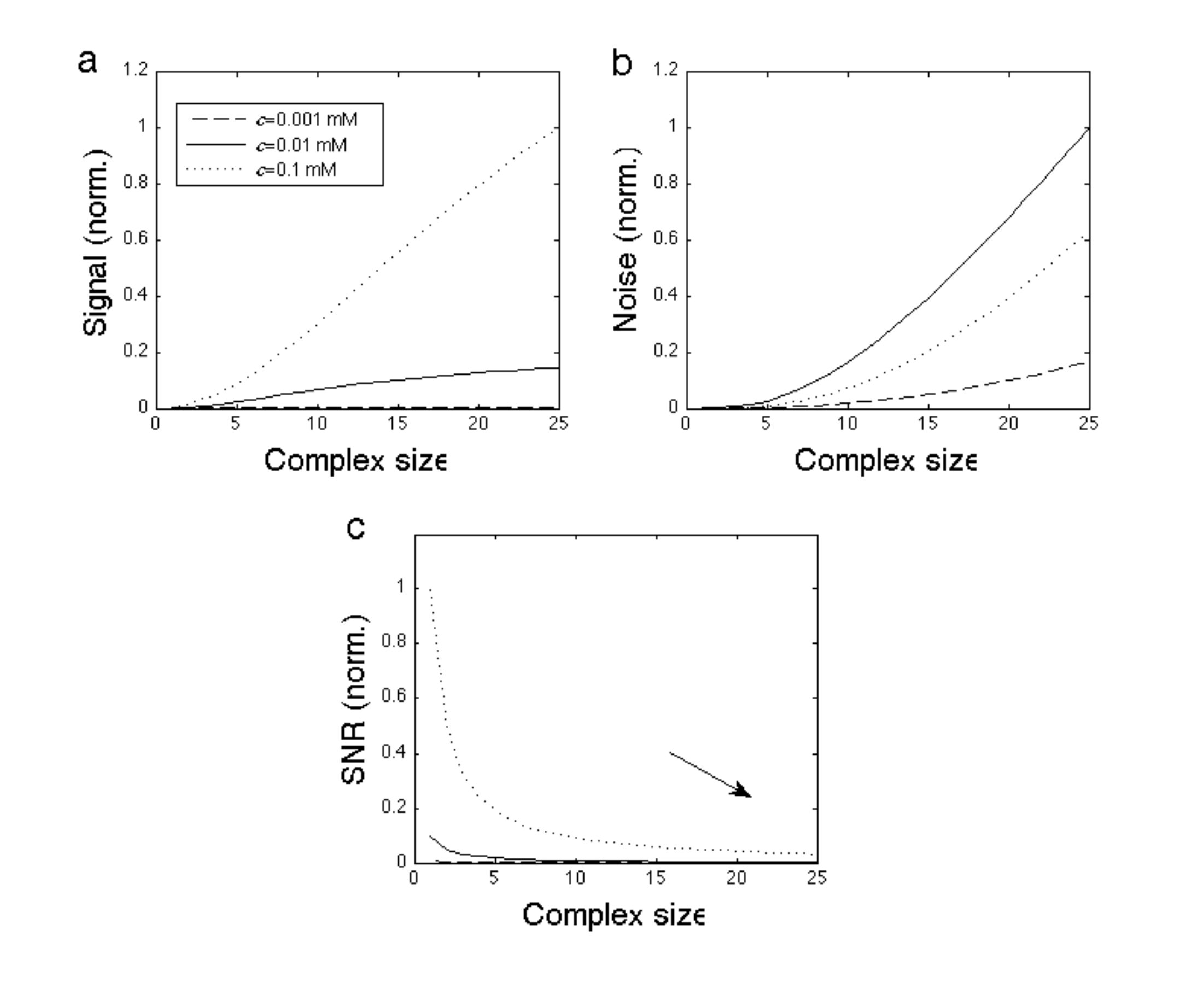}
\caption{\label{fig6} 
Numerical confirmation of scaling behavior for extrinsic ligand noise. (a) Signal, (b) Noise,
and (c) SNR from Fig.~\ref{fig5} as a function of complex size for three different MeAsp 
concentrations $0.001$, $0.01$, and $0.1$ mM. Curves in each panel are normalized by their 
maximal value.
}
\end{figure}

In order to consider the full non-linear activity of the receptors, we numerically integrate 
Eqs.~\eqref{Num2} and \eqref{Denom2} with $m$ set to $\bar m$ and $P(c)$ instead of $P(c,m)$.
Fig.~\ref{fig5} shows contour plots of the Signal, Noise, and SNR 
as a function of complex size and  ligand concentration. In Fig.~\ref{fig6}, the scaling behavior is 
confirmed by plotting the three quantities for three different ligand concentrations as a function of 
complex size.

\ \\

\subsection{\label{sec4b}Ligand and methylation noise} 
Next, we include additional fluctuations in the receptor methylation concentration, as explicitly described by
Eqs. (\ref{Num2}), (\ref{Denom2}), and (\ref{PcM}). The variance of the methylation level is given by
\begin{eqnarray}
\langle(\delta m)^2\rangle&=&\frac{1}{\beta(3-2\bar A)a^3}\nonumber\\
&&\!\!\!\!\!\!\!\!\!\!\!\!\!\!\!\!\!\!\!\!\!\!\!+\frac{k_R(3-2\bar A)(1-\bar A)
(\Delta n N)^2}
{\beta[k_d+k_R(3-2\bar A)(1-\bar A)\beta]ca^3}\label{sM}
\end{eqnarray} 
with $\beta=-\partial F/\partial m$. 
The first term on the right-hand side of Eq.~\eqref{sM} represents the
intrinsic methylation noise, which is independent of complex size consistent with the amplified 
version of Eq.~\eqref{dcMD}. The reason for this $N$-independence is that a large complex has
more enzymes bound to the receptors and hence suffers from larger noise than a small complex. 
However, a large complex also has an increased relaxation rate $k_R+3k_B\bar A^2$, 
restoring the average methylation level more quickly. The second term on the right-hand side of 
Eq.~\eqref{sM} is the ligand-induced methylation noise with its characteristic $N^2$-dependence 
due to amplification by the receptors in the complex.

Furthermore, the covariance between the ligand concentration and methylation level is given by
\begin{equation}
\langle(\delta c)(\delta m)\rangle=\frac{k_R(3-2\bar A)(1-\bar A)\Delta nN}
{k_d+k_R(3-2\bar A)(1-\bar A)\beta}.\label{scM}
\end{equation}
Together with the variances, this equation allows the calculation of the previously mentioned correlation 
coefficient $\rho$. Its value is zero if fluctuations in methylation level are independent of fluctuations in 
ligand concentration, and one if there are no ligand-independent fluctuations in the
methylation level. In our model, the correlation coefficient turns out to be rather small, {\it i.e.} no larger than
0.0001 for the parameter values used.

\begin{figure}[t]
\includegraphics[width=9cm,angle=0]{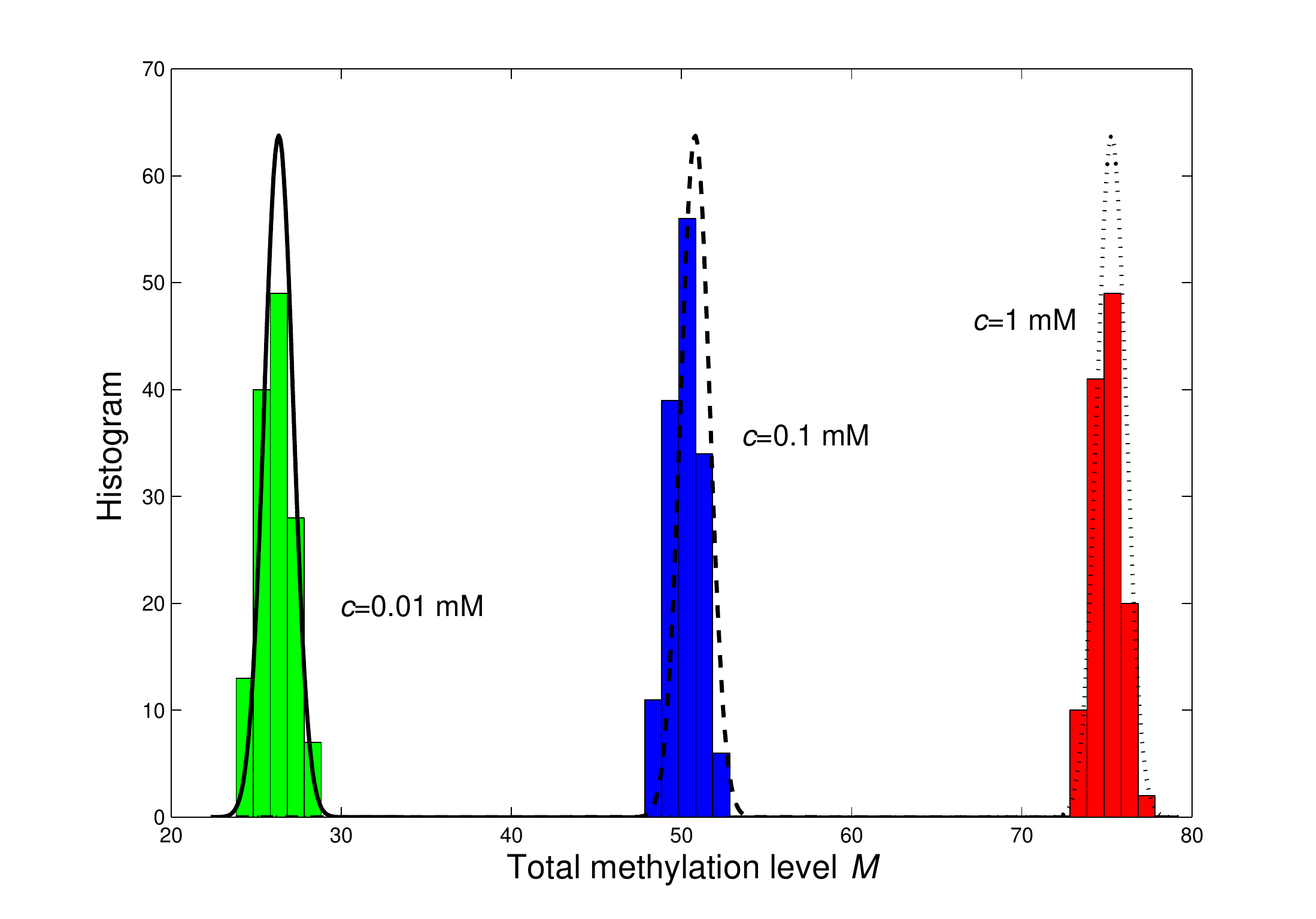}
\caption{\label{fig7} 
(color online) Comparison of the intrinsic methylation noise from the small-noise approximation (van Kampen expansion) with 
Gillespie simulations of the Master equation. Shown are histograms of methylation level from simulations
for external ligand concentrations $c=0.01$ mM (green/left), $c=0.1$ mM (blue/middle), and $c=1$ mM (red/right). 
Also shown are corresponding Normal distributions with variance given by Eq.~\eqref{sM2}, drawn with solid, dashed, 
and dotted lines, respectively.
}
\end{figure}

To check the validity of the small-noise approximation, we compare 
the intrinsic methylation noise from the analytical calculation (first term on the 
right-hand side of Eq.~\eqref{sM}) with simulations of the Master equation using the 
exact Gillespie algorithm \cite{gil77}. (Note the methylation noise is significantly 
larger than the ligand noise and hence is used for this test.) Specifically, the algorithm requires two random numbers. 
The first determines whether to methylate the complex with rate $k_R[1-A(M)]$ or whether to
demethylate with rate $k_BA(M)^3$, where $M=a^3m$ is the current methylation level (the dependence
on the constant external ligand concentration $\bar c$ is not shown). 
The second, $R$, is needed to correctly  increment the simulation time. $R$ is chosen 
with a uniform probability on the interval $[0, 1]$, and the 
time is increased according to $\delta t = 1/\{[k_R(1-A(M)) + k_BA(M)^3]\ln(1/R)\}$. 
Using parameters from appendix \ref{secB} and concentrations ranging from $10^{-3}$ to $1$ mM, 
Fig.~\ref{fig7} shows indeed that the two approaches deliver very similar distributions for 
the methylation level.

We next consider how the cell's Signal and the Noise scale with complex size. First, in the linear activity
regime the Signal is again given by Eq.~\eqref{Num11}. Analoguous to the case with ligand noise only, the 
receptor complex is able to fully amplify the stimulus. (The only difference is that the Signal is 
saturated by the stimulus or noise at smaller complex sizes due to the noise contribution from 
methylation/demethylation.) In contrast, the Noise has a new regime in the presence of additional noise 
from methylation/demethylation
\begin{eqnarray}
<\!\!<\delta A_\text{T}^2>\!\!>_N&\propto&N\left[\langle(\delta A)^2\rangle_m+\langle(\delta A)^2\rangle_c\right]\nonumber\\
&\propto&N(const+N^2).\label{Denom22}
\end{eqnarray}
For small complex sizes, the ligand-induced $N$-dependent activity noise can be neglected 
with respect to the constant contribution from methylation/demethylation, leading to scaling 
with $N$. For larger complex sizes, the Noise scales as $N^3$. The two regimes lead to an optimal
complex size, since the SNR, now given by 
\begin{equation}
\text{SNR}\propto N_T\frac{N}{const+N^2},\label{SNR22}
\end{equation}
first increases proportional to $N$ and then decreases as $N^{-1}$. Hence, the intrinsic noise from methylation/demethylation 
introduces a noise floor, below which it is advantageous for the cell to increase the complex size. However, 
once amplified ligand noise becomes comparable to the noise floor for large complexes, the Noise increases more 
rapidly than the Signal with further increasing complex size.

To address the behavior for the non-linear activity, Figs.~\ref{fig8} and \ref{fig9} show the results from the numerical 
evaluation of the double integrals in Eqs. (\ref{Num2}) and (\ref{Denom2}), confirming our analysis of the scaling. 
For large complex sizes, deviations from the linear regime can be observed. The shape of the optimal complex-size 
curve as a function of ligand concentration is ultimately determined
by the functional dependence of $\Delta n$ on $\bar c$ (Eq.~\eqref{Dn}), which describes the sensitivity of the
receptor occupancy to changes in ligand concentration.

\subsection{Methylation noise}
Finally, we consider the case of only intrinsic noise from fluctuations in the methylation level. 
Eq.~\eqref{PcM} effectively reduces to the Normal distribution 
$P(m)=1/\sqrt{2\pi\langle(\delta m)^2\rangle}\exp\{-(m-\bar m)^2/[2\langle (\delta m)^2\rangle]\}$ 
with average methylation level $\bar m$ and variance
\begin{equation}
\langle (\delta m)^2\rangle=\frac{1}{\beta(3-2\bar A)a^3},\label{sM2} 
\end{equation}
corresponding to the first term in Eq.~\eqref{sM}. In Eqs. (\ref{Num2}) and (\ref{Denom2}), the ligand concentration 
is set to the average value $\bar c$.

Similar to the previous two cases, the Signal behaves as Eq.~\eqref{Num11}, i.e. scales as $N^2$ 
due to stimulus amplification in the linear activity regime. (However, since the methylation/demethylation noise
is independent of complex size, only the stimulus and not the noise can saturate the 
Signal for large complex sizes.) The Noise scales as $N$ from the prefactor in
Eq.~\eqref{Denom2} since the methylation/demethylation noise is independent of complex size. 
As a result, the SNR is proportional to the complex size. Hence, large complexes are always 
better for signaling than small complexes (provided the amplified stimulus is not saturating the Signal). 
This analysis is confirmed by numerical integration in Figs.~\ref{fig10} and \ref{fig11}. 

In summary, intrinsic and extrinsic noise have profoundly different effects on sensing. Only in the presence
of both does an optimal complex size emerge. This result is intuitively clear. Amplification of extrinsic noise
is worse than amplification of the stimulus as receptor complexes behave incoherently for noise amplification
and become fewer in number for increasing complex size.
However, amplification of extrinsic noise is acceptable as long as it stays below the intrinsic noise and 
does not saturate receptor activity. Only when the amplified extrinsic noise becomes larger than the intrinsic 
noise does its effect become detrimental.

\section{\label{sec5} Discussion}

Here we investigated the conditions under which an optimal receptor complex size emerges in order
to provide a potential explanation for the observed receptor cooperativity (Fig.~\ref{fig2}) 
and complexes sizes as imaged by high-resolution microscopy \cite{gre09}. 
Specifically, we considered the signal-to-noise ratio (SNR), {\it i.e.} the ratio between the 
Signal in response to a small stimulus in ligand concentration and the Noise from random 
fluctuations of the activity, based on all the receptor complexes in a cell. 
Using the MWC model for signaling by receptor complexes, we 
include amplification of both stimulus and extrinsic ligand noise, as well as the 
effects of intrinsic noise from random receptor methylation/demethylation. 
Also included are correlations between the ligand and methylation/demethylation noise, since
fluctuations in ligand concentration can induce an adaptational response and hence fluctuations
in methylation level. Note, however, that very slow fluctuations in ligand concentration are fully 
removed by perfect adaptation.

\begin{figure}[t]
\includegraphics[width=9cm,angle=0]{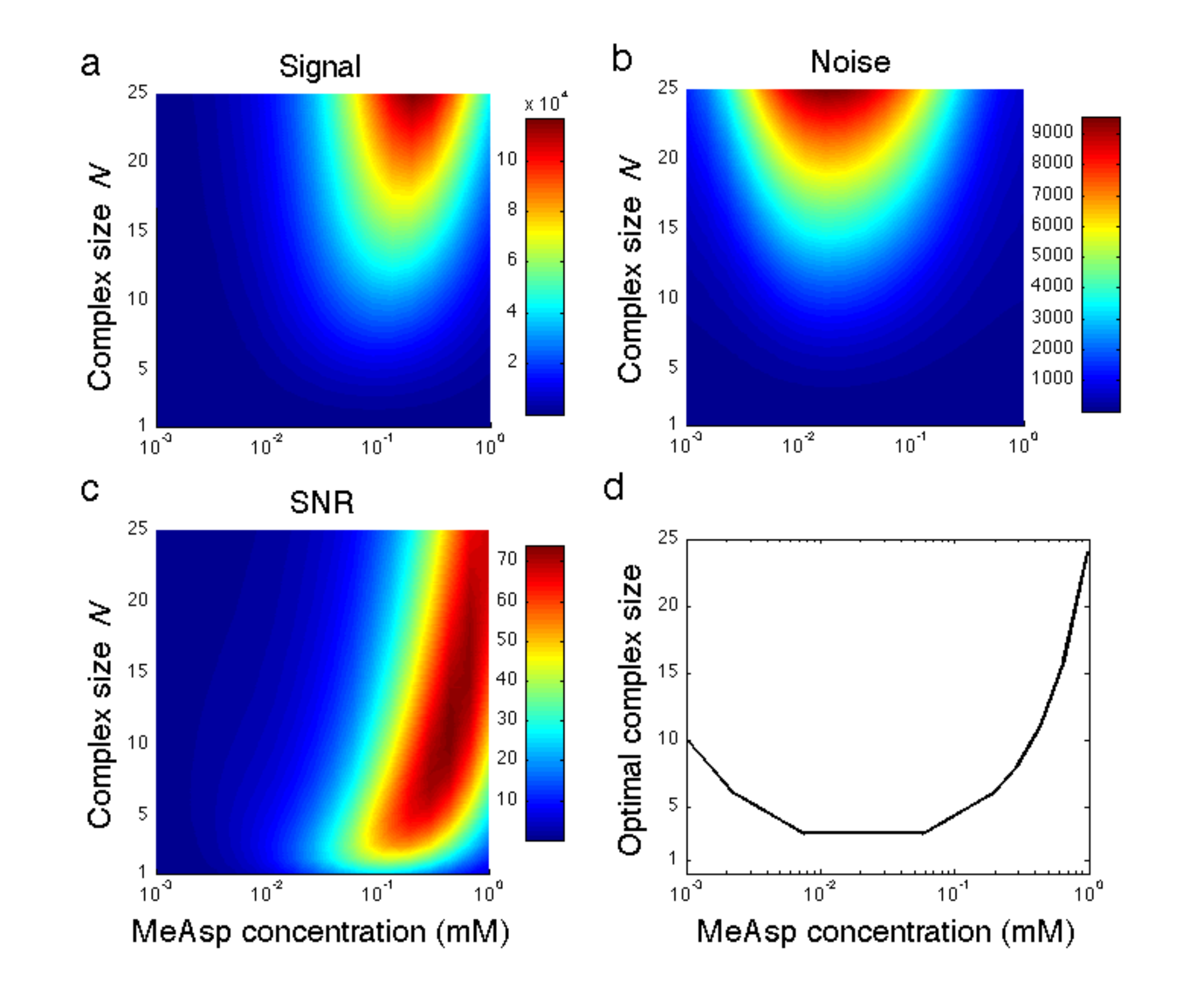}
\caption{\label{fig8} 
(color online) Optimal complex size in the presence of both extrinsic ligand and intrinsic methylation noise. (a) 
Signal, (b) Noise, and (c) SNR as a function of ligand concentration and 
complex size of Tar receptors. 
(d) There is a maximum of the SNR at complex sizes larger than one receptor.  
Parameter values are given in appendix \ref{secB}.}
\end{figure}

\begin{figure}[t]
\includegraphics[width=9cm,angle=0]{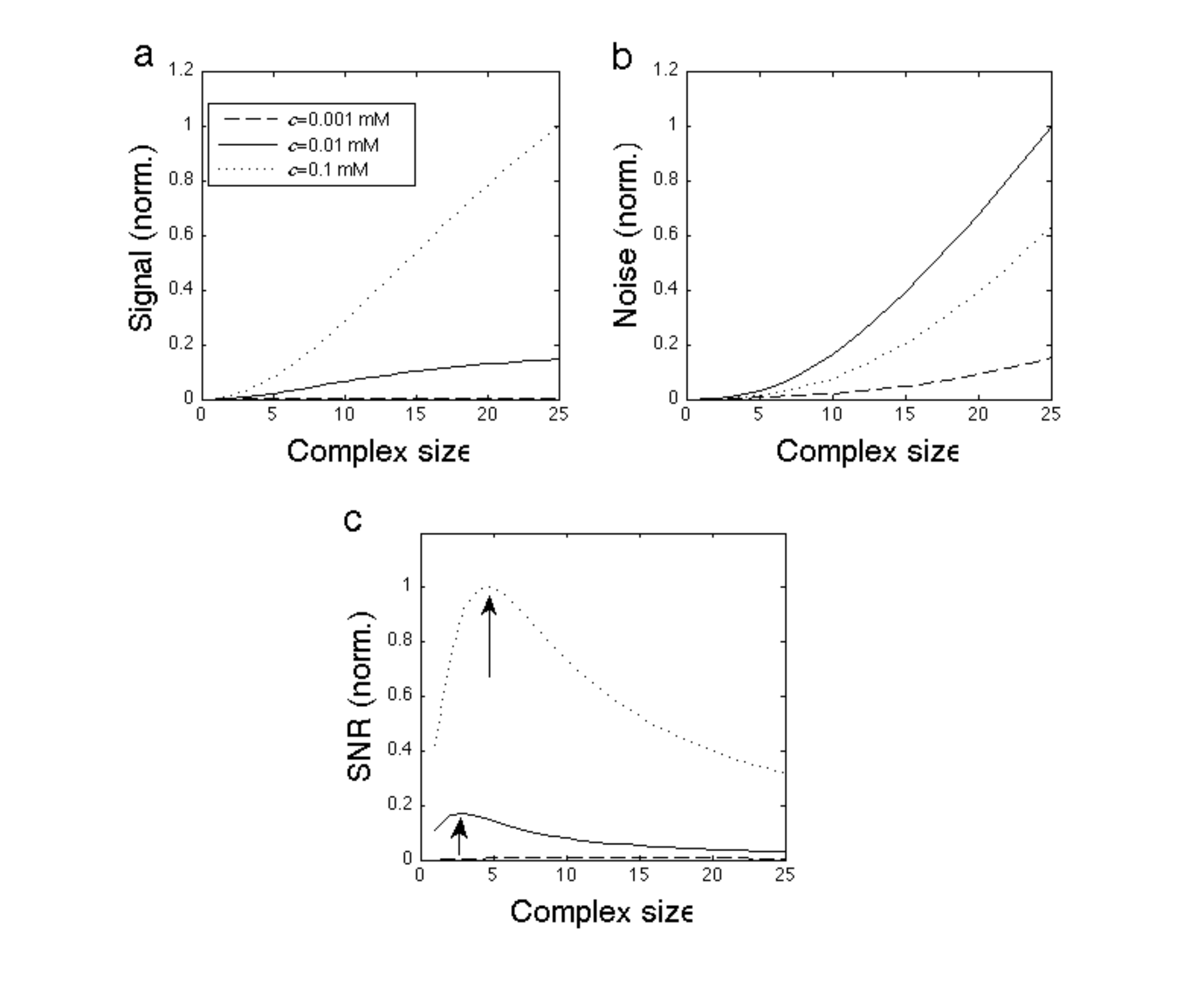}
\caption{\label{fig9} 
Numerical confirmation of scaling behavior in the presence of both extrinsic ligand and intrinsic methylation noise. 
(a) Signal, (b) Noise, and (c) SNR from Fig.~\ref{fig8} as a function of 
complex size for three different MeAsp concentrations $0.001$, $0.01$, and $0.1$ mM. Curves in 
each panel are normalized by their maximal value.}
\end{figure}

\begin{figure}[t]
\includegraphics[width=9cm,angle=0]{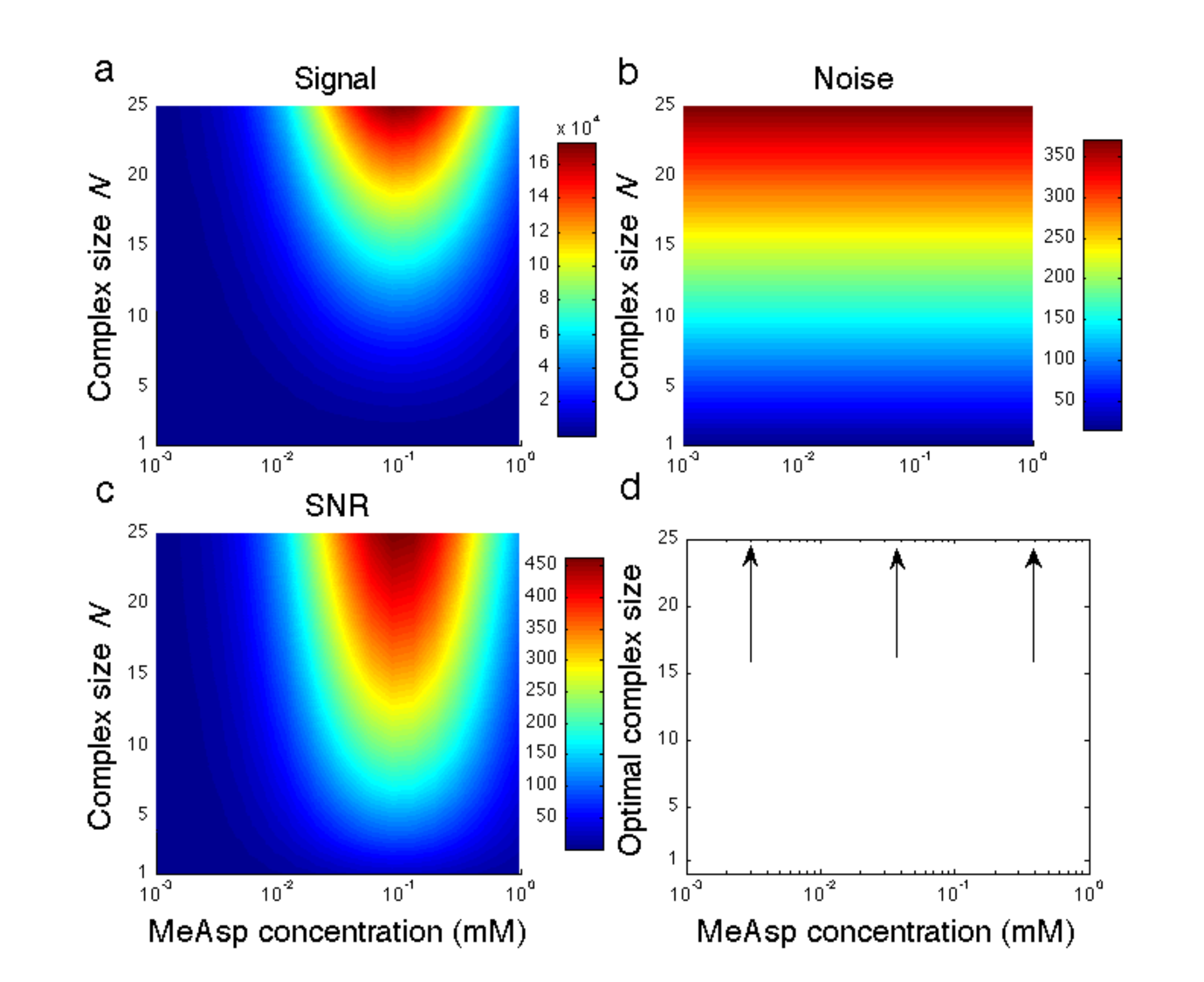}
\caption{\label{fig10} 
(color online) Absence of optimal complex size for intrinsic methylation noise only. (a) 
Signal, (b) Noise, and (c) SNR as a function of ligand concentration and 
complex size of Tar receptors. 
(d) There is no maximum of the SNR since the SNR keeps increasing for increasing complex size. 
Parameter values are given in appendix \ref{secB}.}
\end{figure}

\begin{figure}[t]
\includegraphics[width=9cm,angle=0]{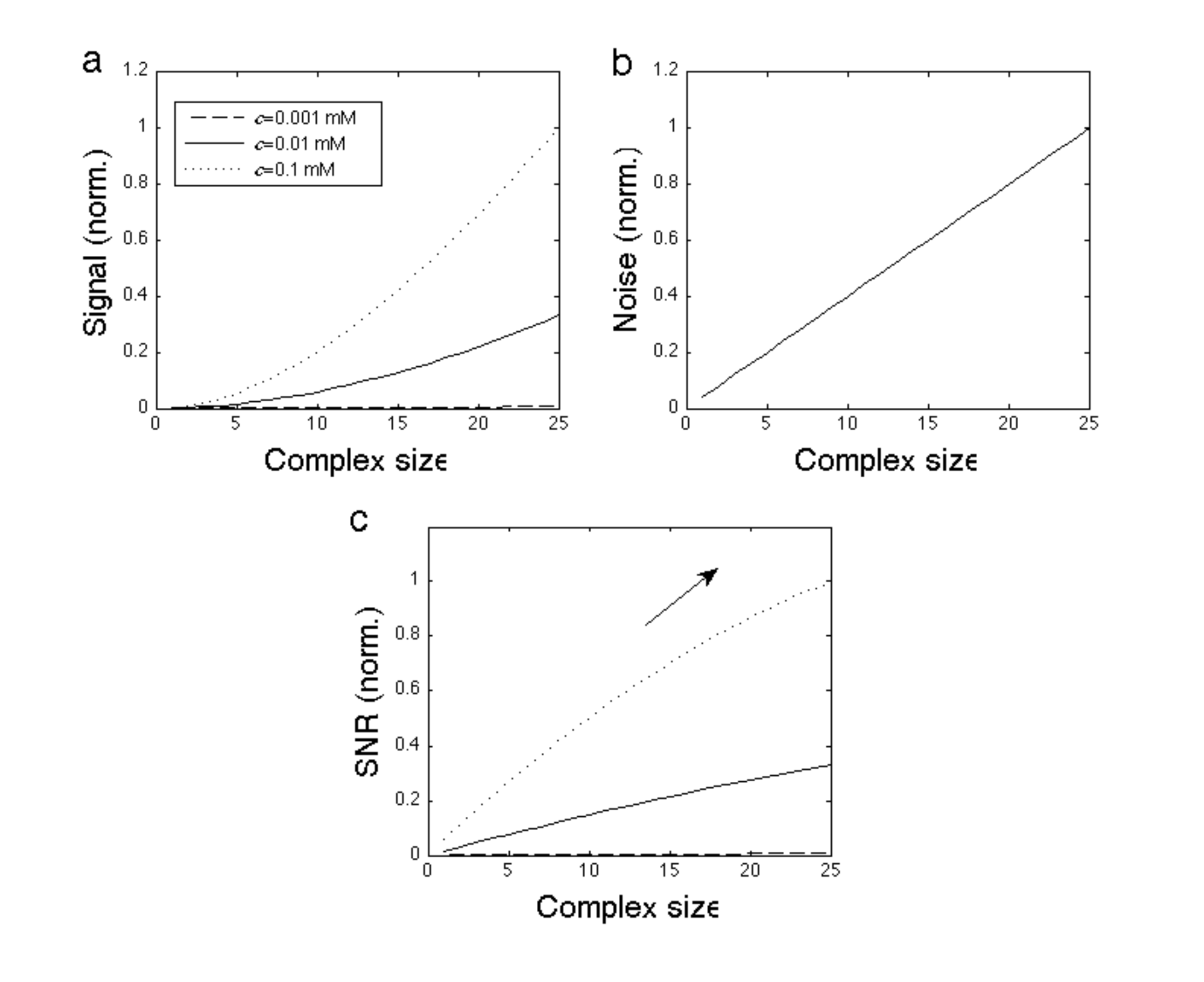}
\caption{\label{fig11} 
Numerical confirmation of scaling behavior for intrinsic methylation noise. 
(a) Signal, (b) Noise, and (c) SNR from Fig.~\ref{fig10} as a function of 
complex size for three different MeAsp concentrations $0.001$, $0.01$, and $0.1$ mM. Curves in 
each panel are normalized by their maximal value.}
\end{figure}

By setting up the Master equation and applying the small-noise approximation, we found that only 
including extrinsic ligand noise leads to a decrease in SNR with complex size
as the Noise inceases more rapidly with complex size than the Signal. When instead only considering 
intrinsic noise, there is no penalty for the cell to make larger and larger complexes, and
the SNR increases with increasing complex size. However, including both noise sources introduces 
a complex size-independent noise floor from the intrinsic noise. Hence, an increase in complex size 
is beneficial until the amplified and hence size-dependent extrinsic noise increases beyond the noise floor. 
An optimal complex size for complexes with more than one receptor is the consequence (Figs.~\ref{fig8} 
and \ref{fig9}).

Extrinsic ligand noise results from the random binding, as well as rebinding of previously 
measured ligand molecules. Importantly, the latter contribution depends on the 
distribution of the receptors. In particular,  the smaller the complex-complex proximity in 
clusters the larger the increase in uncertainty in sensing ligand concentration. 
For ligand diffusion in aqueous solution, the effect of rebinding is generally negligible
(appendix \ref{secC}).
A ligand molecule just released from a receptor is quickly removed by diffusion, hence preventing 
it from rebinding. However, diffusion can be much slower in biologically relevant circumstances.
For instance, in {\it E. coil} chemoreceptors are localized in the inner membrane, which is
surrounded by the dense, viscous periplasm, separating the inner and outer cell membranes.
Here, the ligand diffusion constant can be a thousand times smaller \cite{bra86} and
rebinding of previously bound ligand molecules could be significant.

As demonstrated recently, the detrimental effect of ligand rebinding can be reduced by 
internalization of ligand-bound receptors, such as frequently occurs in 
eukaryotic cells \cite{aqu10}. Receptor internalization effectively turns the cell into an
absorber of ligand molecules, which increases the accuracy of sensing \cite{end08b,aqu10}.
While chemoreceptors in bacteria are not internalized, transporters for uptake of sugars and amino acids 
could colocalize with the receptors to simulate the effect of internalization. Specifically, the uptake
of sugars and amino acids is mainly conducted by periplasmic permeases, which are ABC-like transporters 
\cite{hed94}. The best studied permease is the maltose system in {\it E. coli}. Maltose enters the 
outer membrane through the LamB pores. Subsequently, it is sensed by either directly binding the Tar 
receptor, or maltose-binding protein MalE, which is then either bound by the Tar receptor for sensing 
or by the permease for transport of maltose into the cell. Additional work will be required to 
better understand the role of the periplasm in the accuracy of sensing.

While our model is able to explain the observed complex sizes from FRET data, there are a 
number of simplifying model assumptions. First, we calculated the Signal and Noise at the 
receptor level, neglecting downstream signaling events such as phosphorylation
and dephosphorylation reactions. However, such reactions are known to be very fast, 
$50-1000\,\text{s}^{-1}$ \cite{cla10},  and hence their noise is quickly averaged out by the motor. 
Second, for our calculation of the noise to be computationally feasible, we assume 
diffusion-limited binding to avoid difficulties with the two different receptor complex states.
Third, fast intrinsic noise from random receptor switching between its two activity states 
is assumed to be averaged out and hence is also neglected.

A remaining question is if optimization principles hold for cellular subsystems (here receptor
sensing). There might be tradeoffs, potentially leading to suboptimal solutions for parts of the
cell. Furthermore, receptor sensing is not the final cell's output (here swimming), on which 
natural selection may operate. However, as receptors enable cells to gather information about their 
environment and information can only be lost, not gained during signal transduction, it appears to
be a reasonable assumption to conserve this information as much as possible (if energy and resources 
are not limiting). To instead optimize chemotaxis signaling at the level of cell swimming, not only 
swimming up typical gradients, but also staying on top of gradients (at the maximum concentration) \cite{cla05} 
and the dynamics of gradients \cite{cel10} would need to be considered. 

Our model can readily be extended from Tar-only receptor complexes to mixed complexes of multiple 
receptor types \cite{key06}. If ligand binds specifically to one receptor type only, the receptor 
fraction of each type in a mixed complex should be optimal in size with specifics depending
on the ligand dissociation constants only. Furthermore, our work may be applicable to other 
receptors as well. While receptor clustering may be restricted to receptors with high
sensing accuracy, most well-characterized sensory receptors are believed to cluster (or to
oligomerize). These include the eukaryotic B-cell, T-cell, Fc$\gamma$, synaptic, as well as G-protein-coupled 
and ryanodine receptors \cite{ceb10}. Specifically, T-cell receptors form micro-clusters of 7-30 receptors \cite{lil10}. 
Such receptor aggregates are often associated with lipid rafts, islands of specific lipids with particular 
affinity for certain membrane proteins. Interestingly, lipid rafts were found to be small, containing mostly 
about 6-12 proteins \cite{gur09} and were recently even observed in bacteria \cite{lop10}. 
Our work may indicate that this number represents an optimal size 
for signal amplification, where size is restricted by extrinsic and intrinsic noise.

%\begin{figure}[t]
%\includegraphics[width=8cm,angle=0]{Fig1.pdf}
%\caption{\label{fig:fig1} }
%\end{figure}

%\begin{widetext}
%\begin{equation}
%\end{equation}
%\end{widetext}

\begin{acknowledgments}
We thank S. Neumann, V. Sourjik, and N.S. Wingreen for helful discussions and anonymous referees for 
valuable suggestions. G. A. and R.G.E. acknowledge funding 
from the Biotechnology and Biological Sciences Research Council grant BB/G000131/1, and S. T. and R.G.E. 
thank the Centre for Integrative Systems Biology at Imperial College (CISBIC) for financial support.\\
\end{acknowledgments}

\appendix

\section{\label{secA} Derivation of noise terms using $\Omega$ expansion}

In this appendix, we set up the Master equation of the simplified problem of a receptor complex, whose
activity is determined by the external ligand concentration and receptor methylation level. The dynamics
of the latter are determined by adaptation. To solve for the first and second moments of the joint 
probability distribution
from the Master equation, we apply van Kampen's $\Omega$ expansion, where $\Omega$ is the reaction volume, 
allowing one to introduce a large expansion parameter \cite{kam07}. We neglect fast processes such as 
receptor-complex switching between different activity states and ligand-receptor binding/unbinding. 
Note that we use a slighly different notation in this appendix, more suitable for stochastic processes.

\begin{figure}[t]
\includegraphics[width=8cm,angle=0]{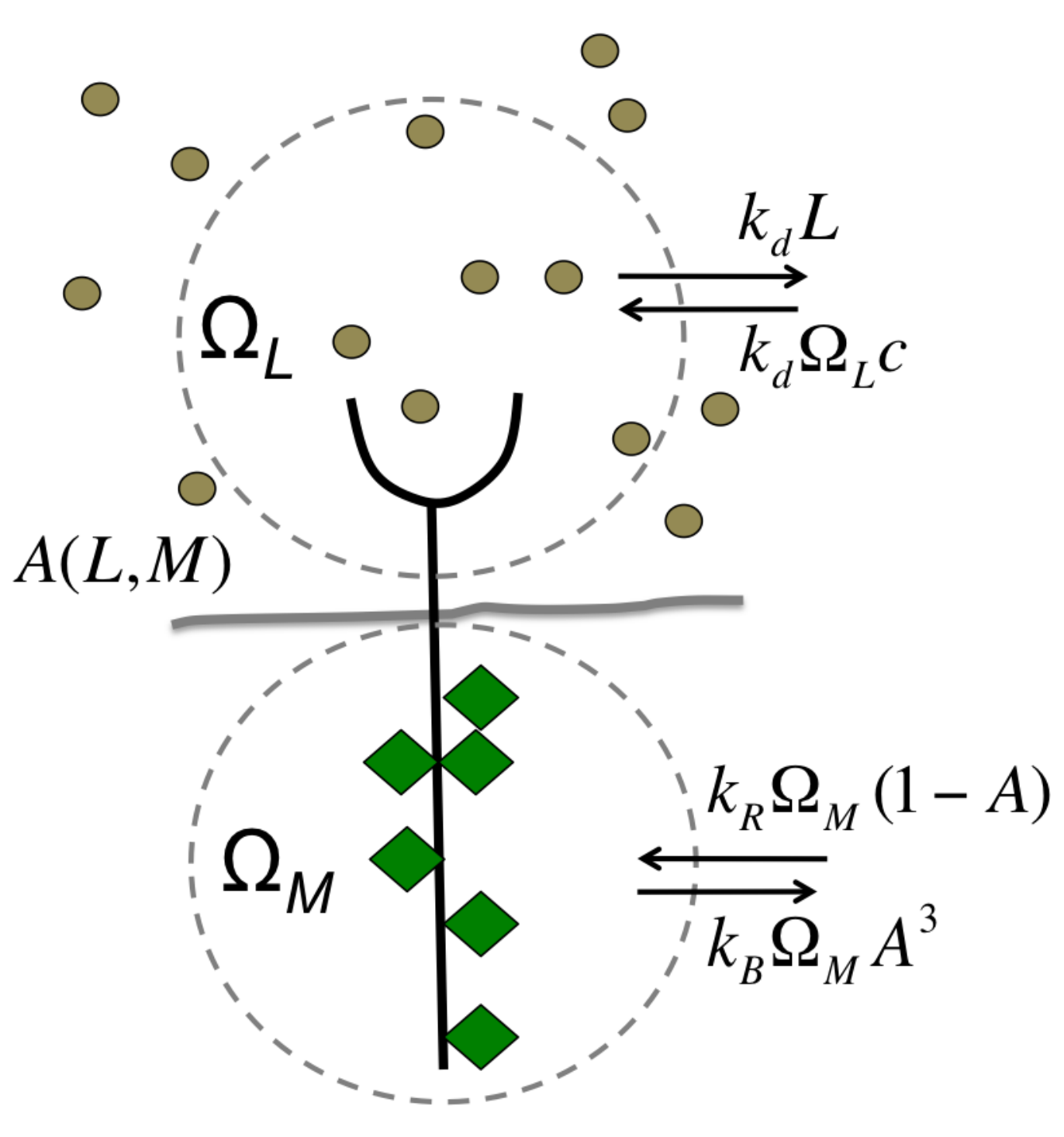}
\caption{\label{fig12} (color online)
Simple model of receptor-complex activity $A(L,M)$ with its dependence on ligand molecules $L$ 
in its vicinity (volume $\Omega_L$) and receptor-methylation level $M$ (in volume $\Omega_M$). Rate constants
$k_d$, $k_R$, and $k_B$ describe ligand diffusion with average ligand concentration $c$, methylation, and 
demethylation, respectively.}
\end{figure}

As shown in Fig.~\ref{fig12} we consider a system composed of two parts. The cell external volume $\Omega_L$
contains the number of ligand molecules $L$ in the vicinity of the receptor complex, which increases from
$L\rightarrow L+1$ with rate $k_d\Omega_L\bar c$ and decreases from $L\rightarrow L-1$ with rate $k_dL$.
The cell internal volume $\Omega_M$ contains the methyl-groups on the receptor, which increase from
$M\rightarrow M+1$ with rate $k_R\Omega_M[1-A(L,M)]$ and decrease from $M\rightarrow M-1$ with 
rate $k_B\Omega_M[A(L,M)]^3$ (cf. macroscopic Eq.~\eqref{dMdt}). The corresponding Master equation
for these one-step processes is given by
\begin{eqnarray}
&&\frac{\partial P(L,M,t)}{\partial t}=k_d\Omega_LcP(L-1,M,t)\nonumber\\
&&+k_d(L+1)P(L+1,M,t)\nonumber\\
&&+k_R[1-A(L,M-1)]P(L,M-1,t)\nonumber\\
&&+k_B[A(L,M+1)]^3P(L,M+1,t)\nonumber\\
&&-\{k_d(L+\Omega_Lc)+k_R[1-A(L,M)]\nonumber\\
&&+k_B[A(L,M)]^3\}P(L,M,t).
\end{eqnarray}
We now define the following separation of $L$ and $M$ into macroscopic parts $c$ and $m$ of respective sizes $\Omega_L$ 
and $\Omega_M$, and fluctuating parts $\zeta$ and $\eta$ of respective sizes $\Omega_L^{1/2}$ and $\Omega_M^{1/2}$,
\begin{subequations}
\begin{align}
L(t)=\Omega_Lc(t)+\Omega_L^{1/2}\zeta(t)\label{L}\\
M(t)=\Omega_Mm(t)+\Omega_M^{1/2}\eta(t).\label{M}
\end{align}
\end{subequations}
We also define the step operators
\begin{subequations}
\begin{align}
\mathbb{E}_L^{+1}f(L)=f(L+1)\\
\mathbb{E}_L^{-1}f(L)=f(L-1)\\
\mathbb{E}_M^{+1}f(M)=f(M+1)\\
\mathbb{E}_M^{-1}f(M)=f(M-1)
\end{align}
\end{subequations}
for any arbitrary function $f(...)$. Using Eqs. \ref{L} and \ref{M}, in the limits of large $\Omega_L$ 
and $\Omega_M$, the step operators adopt the differential form \cite{kam07}
\begin{subequations}
\begin{align}
\mathbb{E}_L^{\pm1}=1\pm\Omega_L^{-1/2}\frac{\partial}{\partial \zeta}+\frac{1}{2}\Omega_L^{-1}\frac{\partial^2}{\partial\zeta^2}\pm....\\
\mathbb{E}_M^{\pm1}=1\pm\Omega_M^{-1/2}\frac{\partial}{\partial \eta}+\frac{1}{2}\Omega_L^{-1}\frac{\partial^2}{\partial\eta^2}\pm....
\end{align}
\end{subequations}
with higher-order terms neglected. Transforming from the old variables $L$ and $M$ to the new variables 
$\zeta$ and $\eta$, we have the relations
\begin{subequations}
\begin{align}
P(L,M,t)\rightarrow\Pi(\zeta,\eta,t)\\
\Omega_L^{1/2}\frac{\partial}{\partial L}P(L,M,t)=\frac{\partial}{\partial \zeta}\Pi(\zeta,\eta,t)\\
\Omega_M^{1/2}\frac{\partial}{\partial M}P(L,M,t)=\frac{\partial}{\partial \eta}\Pi(\zeta,\eta,t).
\end{align}
\end{subequations}
With the above relations, we transform the Master equation, now written with step operators,
\begin{eqnarray}
&&\frac{\partial P(L,M,t)}{\partial t}=k_d\Omega_Lc(\mathbb{E}_L^{-1}-1)P(L,M,t)\nonumber\\
&&\qquad+k_d(\mathbb{E}_L^{+1}-1)LP(L,M,t)\nonumber\\
&&+k_R\Omega_M(\mathbb{E}_M^{-1}-1)[1-A(L,M)]P(L,M,t)\nonumber\\
&&+k_B\Omega_M(\mathbb{E}_M^{+1}-1)[A(L,M)]^3P(L,M,t)
\end{eqnarray}
into 
\begin{eqnarray}
&&\frac{\partial \Pi}{\partial t}-\Omega_L^{1/2}\frac{dc}{dt}\frac{\partial\Pi}{\partial\zeta}
-\Omega_M^{1/2}\frac{dm}{dt}\frac{\partial\Pi}{\partial\eta}\nonumber\\
&=&\!\!\!k_d\Omega_L^{1/2}c\!\!\left[-\frac{\partial}{\partial\zeta}
+\frac{1}{2}\Omega_L^{-1/2}\frac{\partial^2}{\partial\zeta^2}\right]\!\!\Pi(\zeta,\eta,t)\nonumber\\
&+&\!\!\!k_d\Omega_L^{1/2}\left[\frac{\partial}{\partial\zeta}+\frac{1}{2}\Omega_L^{-1/2}\frac{\partial^2}{\partial\zeta^2}\right]
(c+\Omega_L^{-1/2}\zeta)\Pi(\zeta,\eta,t)\nonumber\\
&+&\!\!\!k_R\Omega_M^{1/2}\left[-\frac{\partial}{\partial\eta}+
\frac{1}{2}\Omega_M^{-1/2}\frac{\partial^2}{\partial\eta^2}\right]\nonumber\\
&&\qquad\qquad\cdot[1-A(c,\zeta,m,\eta)]\Pi(\zeta,\eta,t)\nonumber\\
&+&\!\!\!k_B\Omega_M^{1/2}\left[\frac{\partial}{\partial\eta}+
\frac{1}{2}\Omega_M^{-1/2}\frac{\partial^2}{\partial\eta^2}\right]\nonumber\\
&&\qquad\qquad\cdot[A(c,\zeta,m,\eta)]^3\Pi(\zeta,\eta,t).\label{dPidt}
\end{eqnarray}
Next we expand the receptor activity to extract its $\Omega$ dependencies using 
$(A+\delta A)^3=A^3+3A^2\delta A +O(\delta A^2)$ with $\delta A$ a small deviation from activity $A$, and
\begin{equation}
A(c,\zeta,m,\eta)\approx A(c,m)+\frac{\partial A}{\partial m}\Omega_M^{-1/2}\eta+
\frac{\partial A}{\partial c}\Omega_L^{-1/2}\zeta+\dots
\end{equation}
Putting everything together, the terms proportional to $\Omega_L^{1/2}$ produce the macroscopic equation $dc/dt=0$, which
indicates that the ligand concentration $c$ is already at steady state, and the terms proportional to $\Omega_M^{1/2}$ produce
$dm/dt=k_R(1-A)-k_BA^3$ (cf. Eq.~\eqref{dMdt}). 

Importantly, from Eq.~\eqref{dPidt} it is possible to derive equations for the mean value of the fluctuations as well as
for the correlations of these fluctuations. Assuming $\Omega_L=\Omega_M=\Omega$, we first collect all the terms proportional 
to $\Omega^0$ in Eq.~\eqref{dPidt}, which yields 
\begin{eqnarray}
&&\frac{\partial \Pi}{\partial t}=k_d\left[c\frac{\partial^2}{\partial\zeta^2}
+\frac{\partial}{\partial\zeta}\zeta\right]\Pi\nonumber\\
&&+\frac{1}{2}\left\{k_R[1-A(c,m)]
+k_B[A(c,m)]^3\right\}\frac{\partial^2\Pi}{\partial\eta^2}\nonumber\\
&&+k_R\left\{\frac{\partial A}{\partial m}+\left[\frac{\partial A}{\partial c}\zeta+
\frac{\partial A}{\partial m}\eta\right]\frac{\partial}{\partial\eta}\right\}\Pi\nonumber\\
&&\!\!\!\!\!\!\!\!\!\!+3k_B[A(c,m)]^2\left\{\frac{\partial A}{\partial m}+\left[\frac{\partial A}{\partial c}\zeta+
\frac{\partial A}{\partial m}\eta\right]\frac{\partial}{\partial\eta}\right\}\Pi.\label{Omega0}
\end{eqnarray}
Next, multiplying by $\zeta$ and integrating over $\zeta$ and $\eta$ (using intergation by parts), 
produces
\begin{equation}
\frac{\partial \langle\zeta\rangle}{\partial t}=-k_d\langle\zeta\rangle.
\end{equation}
The analoguous procedure for $\eta$ produces
\begin{equation}
\frac{\partial \langle\eta\rangle}{\partial t}=
-(k_R+3k_BA^2)\left[\frac{\partial A}{\partial m}\langle\eta\rangle+\frac{\partial A}{\partial c}\langle\zeta\rangle\right].
\end{equation}
Furthermore, multiplying Eq.~\eqref{Omega0} by $\zeta^2$, $\eta^2$, and $\zeta\eta$ with subsequent 
integration yields, respectively,
\begin{eqnarray}
\frac{\partial\langle\zeta^2\rangle}{\partial t}&=&-2k_d\langle\zeta^2\rangle+2k_dc\\
\frac{\partial\langle\eta^2\rangle}{\partial t}&=&k_R(1-A)+k_BA^3\nonumber\\
&&-2\left(k_R+3k_BA^2\right)\frac{\partial A}{\partial m}\langle\eta^2\rangle\nonumber\\
&&-2(k_R+3k_BA^2)\frac{\partial A}{\partial c}\langle\zeta\eta\rangle\\
\frac{\partial\langle\zeta\eta\rangle}{\partial t}&=&
-\left[k_d+\left(k_R+3k_BA^2\right)\frac{\partial A}{\partial m}\right]\langle\zeta\eta\rangle\nonumber\\
&&-(k_R+3k_BA^2)\frac{\partial A}{\partial c}\langle\zeta^2\rangle.
\end{eqnarray}
At steady state, we finally obtain
\begin{eqnarray}
\langle\zeta^2\rangle_s&=&c\label{zeta2}\\
\langle\eta^2\rangle_s&=&\frac{1}{\beta(3-2\bar A)}\nonumber\\
&&\!\!\!\!\!\!\!\!\!\!\!\!\!\!+\frac{k_R(3-2\bar A)(1-\bar A)(\Delta n N)^2}{[k_d+
k_R(3-2\bar A)(1-\bar A)\beta]c\beta}\label{eta2}\\
\langle\zeta\eta\rangle_s&=&\frac{k_R(3-2\bar A)(1-\bar A)\Delta nN}{k_d+k_R(3-2\bar A)(1-\bar A)\beta}\label{zetaeta}
\end{eqnarray}
with their characteristic $N$-dependencies, $\bar A$ the adapted steady-state activity, and 
$\Delta n$ and $\beta$ defined in Sec. \ref{sec2}. 
The corresponding quantities expressed in the original variables, i.e. $\langle(\delta L)^2\rangle$,
$\langle(\delta M)^2\rangle$, and $\langle(\delta L)(\delta M)\rangle$ are produced by multiplication 
of Eqs. \ref{zeta2}-\ref{zetaeta} with $\Omega$. 
For instance, $\langle(\delta L)^2\rangle=\langle L\rangle$, or $\langle(\delta c)^2\rangle=c/\Omega\sim c/(a^3)$ 
with $a$ the dimension of volume $\Omega$, is indicative of a simple Poisson process and describes the 
instantaneous, total fluctuations in ligand concentration.\\

\section{\label{secB} Parameter values}

Here we provide the parameter values used for the results from section \ref{sec4},
presented in Figs.~\ref{fig5}-\ref{fig11}: $a=100$nm (dimension of receptor complex) \cite{bri08,khu08},
$\beta=a^3/2$ (energy contribution per methyl group to receptor free energy) \cite{end08a},
$k_Ra^3=0.1\,\text{s}^{-1}$ and $k_Ba^3=2.2\,\text{s}^{-1}$ (methylation and demethylation rates) \cite{end06,cla10},
$\bar A=1/3$ (resulting adapted complex activity),
$k_f=10^3\,\text{s}^{-1}$ and $k_b=2\cdot 10^3\,\text{s}^{-1}$ (rates for complex switching) \cite{shar04}, 
$D=300\,\mu\text{m}^2/\text{s}$ (diffusion constant for molecules in aqueous solution)\cite{end08b},                       
$N_T=3000$ (number of receptors per cell) \cite{haz08},
$K_D^\text{on}=0.5$mM and
$K_D^\text{off}=0.02$mM (MeAsp dissociation constants 
for Tar in the on and off states) \cite{key06}.\\

\section{\label{secC} Effect of receptor distribution on uncertainty}
To estimate the magnitude of the structure factor $\Phi$ in Eqs. (\ref{dcSR}) and (\ref{dcMD}), and hence the
importance of ligand rebinding to the uncertainty of sensing ligand concentration, we apply 
the following algorithm for uniformely distributing $N_C$ points (complexes) on a sphere. 
The algorithm divides the sphere in $N_C$ parallels and places a point on each parallel at 
positions given by the following equations in spherical coordinates \cite{saf97}
\begin{subequations}
\begin{align}
h_k&=-1+ 2(k-1)/(N_C-1)  \;\;\;\;  1\leq k\leq N_C\\
\theta_k&=\arccos{h_k} \\
\phi_k&=\left(\phi_{k-1} +\frac{3.6}{\sqrt{N_C}} \frac{1}{\sqrt{1-h_k}}\right) (\text{mod}\,2\pi).
\end{align}
\end{subequations}
The number 3.6 in the algorithm can, in principle, be adjusted appropriately for the application 
at hand but derives essentially from best packing algorithms. The simple algorithm adopted here 
distributes the points spirally
around the sphere and gives a good results in accordance
with more sophisticated algorithms based on energy minimization between point charges
on the sphere or best packing criteria, as long as $N_C>100$ and $N_C<12000$. 

To represent the polar receptor cluster, we choose a small sphere of radius 
$R_S=75$nm \cite{khu08}. As a result, the structure factor for the small sphere is given by
$\Phi_S=4.77\cdot10^9\text{m}^{-1}$. In contrast, to represent receptor complexes evenly
distributed over the cell surface, we use a large sphere of radius $R_L=1.2\,\mu$m, which leads
to a smaller structure factor given by $\Phi_L=5.73\cdot10^8\text{m}^{-1}$. Compared to rebinding
to the same receptor, which is proportional to $1/a=10^7$m${}^{-1}$ in Eq.~\eqref{dcSR}, the 
polar receptor cluster can significantly worsen the uncertainty of sensing. 
However, for fast ligand diffusion in aqueous solution, the
rebinding terms are negligible compared to the other contributions to the uncertainty.

\end{document}